\documentclass[reprint, showpacs, groupedaddress
 amsmath,amssymb,
 aps,
 natbib,
]{revtex4-1}

\usepackage{graphicx}
\usepackage{dcolumn}
\usepackage{bm}
\usepackage{hyperref}
\usepackage{amssymb}
\usepackage{mathrsfs}
\usepackage{amsmath,fleqn}

\newcommand{\bea}{\begin{eqnarray}}
\newcommand{\eea}{\end{eqnarray}}
\newcommand{\ba}{\begin{array}}
\newcommand{\ea}{\end{array}}

\newcommand{\be}{\begin{equation}}

\newcommand{\ee}{\end{equation}}
\newcommand{\bt}{\begin{teo}}
\newcommand{\et}{\end{teo}}

\newcommand{\om}{\omega}

\newcommand{\bro}{\beta_{\rm BR}}
\newcommand{\blo}{\beta_{\rm loc}}
\newcommand{\bex}{b}
\newcommand{\biz}{\beta_{\rm IZ}}
\newcommand{\La}{\Lambda}
\newcommand{\linf}{\l_{\infty}}

\begin{document}

\preprint{APS/123-QED}

\title{Dynamical localization in chaotic systems:\\ spectral statistics
and localization measure in kicked rotator \\ as a paradigm
for time-dependent and time-independent systems }

\author{Thanos Manos}
\email{thanos.manos@uni-mb.si}
\affiliation{CAMTP - Center for Applied Mathematics and Theoretical Physics, University of Maribor, Krekova 2, SI-2000 Maribor, Slovenia}
\affiliation{School of Applied Sciences, University of Nova Gorica, Vipavska 11c, SI-5270 Ajdov\v s\v cina, Slovenia, European Union.}

\author{Marko Robnik}
\email{Robnik@uni-mb.si}
\affiliation{%
 CAMTP - Center for Applied Mathematics and Theoretical Physics, University of Maribor, Krekova 2, SI-2000 Maribor, Slovenia, European Union.
}

\date{\today}

\begin{abstract}
We study the kicked rotator in the classically fully chaotic regime using Izrailev's $N$-dimensional model for various $N \le 4000$, which in the limit $N \rightarrow \infty$ tends to the quantized kicked rotator. We do not treat only the case $K=5$ as studied previously, but many different values of the classical kick parameter $5\le K \le 35$, and also many different values of the quantum parameter $k\in [5,60]$. We describe the features of dynamical localization of chaotic eigenstates as a paradigm for other both time-periodic and time-independent (autonomous) fully chaotic or/and mixed type Hamilton systems. We generalize the scaling variable $\La =l_{\infty}/N$ to the case of anomalous diffusion in the classical phase space, by deriving the localization length $l_{\infty}$ for the case of generalized classical
diffusion.  We greatly improve the accuracy and statistical significance of the numerical calculations, giving rise to the following conclusions: (C1) The level spacing distribution of the eigenphases (or quasienergies) is very well described by the Brody distribution, systematically better than by other proposed models, for various Brody exponents $\bro$.  (C2) We study the eigenfunctions of the Floquet operator and characterize their localization properties using the information entropy measure, which after normalization is given by $\blo$ in the interval $[0,1]$. The level repulsion parameters $\bro$ and $\blo$ are almost linearly related, close to the identity line. (C3) We show the existence of a scaling law between $\blo$ and the relative localization length $\La$, now including the regimes of anomalous diffusion. The above findings are important also for chaotic eigenstates in time-independent systems (Batisti\'c and Robnik 2010,2013), where the Brody distribution is confirmed to a very high degree of precision for dynamically localized chaotic eigenstates even in the mixed-type systems (after
separation of regular and chaotic eigenstates).
\end{abstract}

\pacs{05.45.Mt,03.65.-w,05.45.Pq,03.65.Aa}
\keywords{Suggested keywords}
\maketitle


\section{Introduction \label{intro}}

One of the main cornerstones in the development of quantum chaos \cite{Stoe,Haake,Rob1998} is the finding that in classically fully chaotic, ergodic, autonomous Hamilton systems with the purely discrete spectrum the fluctuations of the energy spectrum around its mean behaviour obey the statistical laws described by the Gaussian Random Matrix Theory (RMT) \cite{Mehta,GMW}, provided that we are in the sufficiently deep semiclassical limit. The latter condition means that all relevant classical transport times are smaller than the so-called Heisenberg time, or break time, given by $t_H=2\pi\hbar/\Delta E$, where $h=2\pi\hbar$ is the Planck constant and $\Delta E$ is the mean energy level spacing, such that the mean energy level density is $\rho(E) =1/\Delta E$. This statement is known as the Bohigas -
Giannoni - Schmit (BGS) conjecture and goes back to their pioneering paper in 1984 \cite{BGS}, although some preliminary ideas were published in \cite{Cas}.  Since $\Delta E \propto \hbar^f$, where $f$ is the number of degrees of freedom (= the dimension of the configuration space), we see that for sufficiently small $\hbar$ the stated condition will always be satisfied. Alternatively, fixing the $\hbar$, we can go to high energies such that the classical transport times become smaller than $t_H$. The role of the antiunitary symmetries that classify the statistics in terms of GOE, GUE or GSE (ensembles of RMT) has been elucidated in \cite{RB1986}, see also \cite{Rob1986}, and \cite{Stoe,Haake,Rob1998,Mehta}. The theoretical foundation for the BGS conjecture has been initiated first by Berry \cite{Berry1985}, and later further developed by Richter and Sieber \cite{Sieber}, arriving finally in the almost-final proof proposed by the group of F. Haake \cite{Mueller1,Mueller2,Mueller3,Mueller4}.

On the other hand, if the system is classically integrable, Poisson statistics applies, as is well known and goes back to the work by Berry and Tabor in 1977 (see \cite{Stoe,Haake,Rob1998} and the references therein, and for the recent advances \cite{RobVeb}).

In the mixed type regime, where classical regular regions coexist in the classical phase space with the chaotic regions, being a typical KAM-scenario which is the generic situation, the so-called Principle of Uniform Semiclassical Condensation (of the Wigner functions of the eigenstates; PUSC) applies, based on the ideas by Berry \cite{Berry1977}, and further extended by
Robnik \cite{Rob1998}. Consequently the Berry-Robnik statistics \cite{BR1984,
ProRob1999}  is observed - see also \cite{Rob1998} - again under the same  semiclassical condition stated above requiring that $t_H$ is larger than all classical transport times.

The relevant papers dealing with the mixed type regime after the work \cite{BR1984} are \cite{ProRob1993a,ProRob1993b,ProRob1994a,ProRob1994b,Pro1998a,Pro1998,GroRob1,GroRob2} and the most recent advance was published in \cite{BatRob2010}, while \cite{BatRob2012} is the relevant work in progress. If the couplings
between the regular eigenstates and chaotic eigenstates become important, due to the dynamical tunneling, we can use the ensembles of random matrices that capture these effects \cite{VSRKHG}.  As the tunneling strengths typically decrease exponentially with the inverse effective Planck constant, they rapidly disappear with increasing energy, or by decreasing the value of the Planck constant.

Here it must be emphasized that the analogies between the time-periodic systems (the kicked
rotator) and time-independent systems (like mixed-type billiards) that we are drawing and
studying in this paper refer to {\bf the chaotic eigenstates} only, which means that we have to
conceptually separate the regular and the chaotic eigenstates in each system.
If the semiclassical condition is satisfied, then for the subspectrum of the
chaotic eigenstates we find extendedness and GOE statistics. This should be compared with the
extended states in finite dimensional kicked rotator model for $K \ge 7$, where
the corresponding classical dynamics is fully chaotic.

However, if the semiclassical condition is not satisfied, such that $t_H$ is no longer larger than the relevant classical transport time, like e.g. the diffusion time in fully chaotic but slowly ergodic systems, we find the so-called {\bf dynamical localization} (or {\bf Chirikov localization}) first observed in time-dependent systems (see e.g. \cite{Stoe}), which are the main topics of the present work and will be discussed below in detail, but later on analyzed quite systematically  in autonomous (time-independent) systems by many authors. For an excellent  review see the paper by Prosen \cite{Pro2000} and the references therein. In such a situation it turns out that the Wigner functions of the chaotic eigenstates no longer uniformly occupy the entire classically accessible chaotic region in the classical phase space, but are localized on a proper subset of it. In contradistinction to the tunneling effects, these dynamical localization effects can survive to very high lying eigenstates. Indeed, this has been analyzed with unprecedented precision and statistical significance by Batisti\'c and Robnik \cite{BatRob2010} in case of mixed type systems, and the work is being extended in the analysis of separated regular and chaotic eigenstates \cite{BatRob2012,BatManRob2013}. The most important discovery is that the level spacing distribution of the
dynamically localized chaotic eigenstates is very well described by the Brody distribution, introduced in \cite{Bro1973}, see also \cite{Bro1981},
with the Brody parameter values $\bro$ within the interval $[0,1]$, where $\bro=0$ yields the Poisson distribution in case of the strongest localization, and $\bro=1$ gives the Wigner surmise (2D GOE, as an excellent approximation of the infinite dimensional GOE). To our great surprise the Brody distribution  fits the empirical data much better than the distribution function proposed by F. Izrailev (see \cite{Izr1988,Izr1990} and the references therein)
characterized by the parameter $\biz$.  This is still true also for the improved Izrailev distribution
published in \cite{CIM1991} and recently used in \cite{SIZC2012}.
It is well known that Brody distribution so far has no theoretical foundation, but our empirical results show that we have to consider it seriously thereby being motivated for seeking its physical foundation.

In the present study of the kicked rotator, besides the above mentioned results on the relevance of the Brody distribution,
we go beyond Izrailev's results in that we study not only the case of the
classical kick parameter $K=5$, but for many other $K \in[5,35]$, and many different values of the
quantum parameter $k$, and consider the relevance of the classical diffusion in greater
depth, allowing also for the anomalous diffusion. In so doing we greatly generalize and improve the
evidence for the linear relationship between the information entropy localization measure $\blo$
and $\bro$, and also for the scaling relationship between $\blo$ and the scaling variable $\La$,
which is the theoretical localization length divided by the dimension of the system.

Our work corroborates the view (see \cite{Izr1990} and the references therein, especially \cite{Izr1986,Izr1987,Izr1989}) that time-independent and time-periodic chaotic systems have many properties in common when it comes to the statistical properties of discrete energy spectra and the discrete quasienergy spectra (or eigenphases), respectively. We think that this view can be extended also to quantifying the degree of localization in such systems. Very recent results (Batisti\'c and Robnik 2013) confirm this expectation, and will be published
separately.

The paper is organized as follows: In section \ref{secmod} we introduce the model system, in section \ref{locdiff} we describe the aspects of generalized diffusion in the classical system (standard map) including the accelerator modes and the anomalous diffusion and relate it to the quantum localization properties, deriving the new formula for the localization length. In section \ref{sec2} we define the finite dimensional quantum model system, introduced by Izrailev and study not only the cases of quantum resonance, but also the generic cases. In section \ref{sec3} we define the information entropy localization measure $\blo$, in section \ref{sec4} we study the statistical properties of spectra (eigenphases), in section \ref{sec5} we analyze the relationship between the localization parameter $\blo$ and the spectral level repulsion parameters $\bro$ and $\biz$, and also study the scaling relationship between $\blo$ and the scaling parameter $\La$. In section \ref{sec6} we draw the final conclusions and discuss the results. In Appendix \ref{ap:A} we define and explain the $U$-function of the level spacings and in Appendix \ref{ap:B} we show some additional results on energy level statistics of the quantum kicked rotator.

\section{Introducing the model \label{secmod}}
One of the main models of time-dependent systems is the kicked rotator introduced by Casati, Chirikov, Ford and Izrailev in 1979 \cite{CCFI79}. We introduce it here in detail for the purpose of defining and fixing the variables and the notation. The Hamiltonian function is
\be  \label{KR}
H= \frac{p^2}{2I} + V_0 \,\delta_T(t)\,\cos \theta.
\ee
It is one of the most important paradigms of classical conservative (Hamiltonian) systems in nonlinear dynamics. Here $p$ is the (angular) momentum, $I$ the moment of inertia, $V_0$ is the strength of the periodic kicking, $\theta$ is the (canonically conjugate, rotation) angle, and $\delta_T(t)$ is the periodic
Dirac delta function with period $T$. Since between the kicks the rotation
is free, the Hamilton equations of motion can be immediately integrated,  and thus the dynamics can be reduced to the standard mapping, or so-called Chirikov-Taylor mapping, given by
\be \label{SM1}
p_{n+1} = p_n + V_0 \sin \theta_{n+1},\;\;\; \theta_{n+1} = \theta_n + \frac{T}{I} p_n,
\ee
and introduced in \cite{T69,F72,C79}. Here the quantities $(\theta_n, p_n)$ refer to their values just immediately after the $n$-th kick. Obviously,
by introducing new dimensionless momentum $P_n = p_nT/I$, we get
\be \label{SM2}
P_{n+1} = P_n + K \sin \theta_{n+1},\;\;\; \theta_{n+1} = \theta_n  + P_n,
\ee
where the system is now governed by a single classical {\em dimensionless}
control parameter $K=V_0 T/I$, and the mapping is area preserving.

The quantum kicked rotator (QKR) is the quantized version of Eq.~(\ref{KR}),
namely
\be \label{QKR}
\hat{H} =-\frac{\hbar^2}{2I} \frac{\partial^2}{\partial\theta^2} +
V_0\, \delta_T(t)\,\cos \theta .
\ee
The physics of the QKR  is extremely rich and it is a paradigm of quantum chaos in Floquet (= time-periodic) systems \cite{Izr1990}. It is also relevant for the autonomous Hamilton systems as indicated above. For such a Floquet system the Floquet operator $\hat{F}$ acting on the wavefunctions (probability amplitudes) $\psi(\theta)$, $\theta \in[0,2\pi)$, upon each period (of length $T$) can be written as (see e.g. \cite{Stoe}, Chapter 4)
\be \label{Fop}
\hat{F} = \exp \left( -\frac{iV_0}{\hbar} \cos\theta\right)
\exp\left(-\frac{i\hbar T}{2I}\frac{\partial^2}{\partial\theta^2}\right),
\ee
where now we have obviously two {\em dimensionless} quantum control parameters
\be \label{qpar}
k=\frac{V_0}{\hbar}, \;\;\; \tau= \frac{\hbar T}{I},
\ee
which satisfy the relationship $K = k\tau = V_0 T/I$, $K$ being the classical {\em dimensionless} control parameter of Eq.~(\ref{SM2}). By using the angular momentum eigenfunctions
\be \label{Eigenf}
|n\rangle = a_n(\theta) = \frac{1}{\sqrt{2\pi}} \exp (i\,n\,\theta),
\ee
where $n$ is any integer, we find the matrix elements of $\hat{F}$, namely
\begin{flalign} \label{Fmatrix}
F_{m\,n} = \langle m|\hat{F}|n\rangle = \exp\left(-\frac{i\tau}{2} n^2\right) i^{n-m} J_{n-m} (k),
\end{flalign}
where $J_{\nu}(k)$ is the $\nu$-th order Bessel function. For a wavefunction
$\psi (\theta)$ we shall denote its angular momentum component (Fourier
component) by
\begin{flalign} \label{Fourier} \nonumber
u_n = \langle n|\psi\rangle = \int_0^{2\pi} a_n^{*}(\theta) \psi(\theta)\,d\theta= \\ =\frac{1}{\sqrt{2\pi}} \int_0^{2\pi} \psi(\theta) \exp(-in\theta)\,d\theta.
\end{flalign}
The QKR has very complex dynamics and spectral properties. As the phase space is infinite (cylinder), $p\in (-\infty, +\infty), \theta\in[0,2\pi)$, the spectrum of the eigenphases of $\hat{F}$, denoted by $\phi_n$, or the associated quasienergies $\hbar\om_n= \hbar \phi_n/T$, introduced by Zeldovich  \cite{Zel1966}, can be continuous, or discrete. It is quite well understood that for the resonant values of $\tau$
\be \label{Restau}
\tau = \frac{4\pi r}{q},
\ee
$q$ and $r$ being positive integers without common factor, the spectrum is continuous, as rigorously proven by Izrailev and Shepelyansky \cite{IS1979a,IS1979b,IS1980a,IS1980b}, and the dynamics is (asymptotically) ballistic, meaning that starting from an arbitrary initial state the mean value of the momentum $\langle\hat{p}\rangle$ increases {\em linearly} in time, and the energy of
the system $E= \langle\hat{p}^2\rangle/(2I)$ grows {\em quadratically} without limits. For the special case $q=r=1$ this can be shown elementary. Such behaviour is a purely quantum effect, called the quantum resonance. Also, the regime of quadratic energy growth manifests itself only after very large time, which grows very fast with the value of the integer $q$ from Eq.~(\ref{Restau}), such that for larger $q$ this regime practically cannot be observed.

For generic values of $\tau/(4\pi)$, being irrational number, the spectrum is expected
to be discrete but infinite. But the picture is very complicated. Casati and Guarneri \cite{CG1984}
have proven, that for $\tau/(4\pi)$ sufficiently close to a rational number, there exists
a continuous component in the quasienergy spectrum. So, the absence of dynamical localization
for such cases is expected as well. Without a rigorous proof, we finally believe that
for all other (``good") irrational values of $\tau/(4\pi)$ we indeed have discrete spectrum and
quantum dynamical localization.
In such case the quantum dynamics is almost periodic, and because of the effective finiteness of the relevant set of components $u_n$ and of the basis functions involved, just due to the exponential localization (see below), it is even effectively quasiperiodic (effectively there is a finite number of frequencies), and any initial state returns after some recurrence time arbitrarily close to the initial state. Thus the energy cannot grow indefinitely.

\section{Localization and diffusion properties\label{locdiff}}

In the generic (nonresonant) case we observe in the semiclassical regime of large $k \gg 1$ and in the classically chaotic regime  $K \ge K_{crit} \approx 0.9716...$, the so-called {\em dynamical localization} also called {\em Chirikov localization}: starting from an initial semiclassical wave packet of the width smaller than the localization length, to be precisely defined below, the average energy grows first according to the classical diffusion, but stops after a finite time, i.e. the localization time $t_{\rm loc}$ (physical time divided by the period of kicking $T$, that is the number of kicks, and thus dimensionless), which is derived below.

The asymptotic localized eigenstates are quasistationary, they just oscillate under the action of $\hat{F}$, as the quantum recurrence time is very large. They are very special, as their expansion coefficients in the basis of the angular momentum eigenstates $|n\rangle$ must be highly correlated. In fact, more can be said about these asymptotic eigenstates: they are {\em exponentially localized}. The (dimensionless) localization length in the space of the angular momentum quantum numbers is derived below, and is equal (after introducing some numerical correction factor $\alpha_{\mu}$) to the dimensionless localization time $t_{\rm loc}$  [Eq.~(\ref{finallinf})]. We denote it like in reference \cite{Izr1990} by $l_{\infty}$.  Therefore, an exponentially localized eigenfunction centered at $m$ in the angular momentum space [Eq.~(\ref{Eigenf})] has the following form
\be \label{exploc}
|u_n|^2 \approx \frac{1}{l_{\infty}} \exp\left(-\frac{2|m-n|}{l_{\infty}}\right),
\ee
where $u_n$ is the probability amplitude [Eq.~(\ref{Fourier})] of the localized wavefunction $\psi(\theta)$. The argument leading to $t_{\rm loc}$ in Eq.~(\ref{finallinf}) originates from the
observation of the dynamical localization by Casati et al \cite{CCFI79}, and in particular from \cite{CIS1981}, and and is well explained in \cite{Stoe}, in case of normal diffusion $\mu=1$, whilst for general $\mu$ we give a theoretical argument in this section.

Since the spectrum is discrete we can ask questions about the statistical properties of the spectrum of the quasienergies, or eigenphases. However, since the system is infinite with infinitely many exponentially localized eigenstates, we have infinitely many eigenphases on the interval $[0,2\pi)$, resulting in an infinite level density, and thus all level spacings are zero. Nevertheless, for any finite but arbitrarily large number of eigenstates
$N$ everything is well defined. In the classically fully chaotic regime one would naively expect the applicability of the RMT, in our case the GOE statistics. However, this is not observed. On the contrary, the statistics
is Poissonian, which is the consequence of the finite localization length at any $k$ in the infinite basis of the angular momentum. Following the heuristic argument by Izrailev, we can say that eigenstates can be quasienergetically very close to each other, an almost degenerate pair, although they are located
in the angular momentum space very far from each other and practically do not overlap due to the exponential localization. Therefore, they do not ``feel" each other, they do not interact, in the sense that changing slightly one of them does not change the other one, and thus contribute to the spectrum in a completely uncorrelated way. This results in the Poissonian statistics.

The question arises, where do we see the analogous phenomena predicted by
the RMT and observed in the quantum chaos of time-independent bound systems with discrete spectrum. To see these effects the system must have effectively finite dimension. Truncation of the infinite matrix $F_{mn}$ in Eq.~(\ref{Fmatrix}) in {\em tour de force} is not acceptable, even in the technical case of numerical computations, since after truncation the Floquet operator is no longer unitary.

The only way to obtain a quantum system which shall in this sense correspond to the classical dynamical system [Eqs.~(\ref{KR}), (\ref{SM1}) and (\ref{SM2})] is to introduce a finite $N$-dimensional matrix, which is symmetric unitary, and which in the limit $N\rightarrow\infty$ becomes the infinite dimensional system with the Floquet operator [Eq.~(\ref{Fop})]. The semiclassical limit is $k\rightarrow \infty$ and $\tau\rightarrow 0$, such that $K=k\tau ={\rm constant}$. As it is well known \cite{Izr1990}, for the reasons discussed above, the system behaves very similarly for rational and irrational values of $\tau/(4\pi)$. Such a $N$-dimensional model \cite{Izr1988} will be introduced in the next section \ref{sec2}.

Let us now derive the semiclassical estimate of the localization time $t_{\rm loc}$ and the localization length $l_{\infty}$, both being dimensionless. It turns out that they are equal, as shown in Eq.~(\ref{tloc=linf}). The generalized diffusion process of the standard map [Eq.~(\ref{SM2})] is defined by
\be \label{varp}
\langle(\Delta P)^2\rangle = D_{\mu}(K) n^{\mu},
\ee
where $n$ is the number of iterations (kicks), and the exponent $\mu$ is in the interval $[0,2)$, and all variables $P$, $\theta$ and $K$ are dimensionless. Here $D_{\mu}(K)$ is the {\bf generalized classical diffusion constant}.
In case $\mu=1$ we have the normal diffusion, and $D_1(K)$ is then the normal diffusion constant, whilst in case of anomalous diffusion we observe subdiffusion when $0 < \mu < 1$  or superdiffusion if $1 <\mu \le2$. In case $\mu=2$ we have the ballistic transport which is associated with the presence of accelerator modes (see below).

As the real physical angular momentum $p$ and $P$ are connected by $P=pT/I$ we have for the variance of $p$ the following equation
\be \label{varpreal}
\langle (\Delta p)^2\rangle = \frac{I^2}{T^2} D_{\mu} n^{\mu}.
\ee
Now we argue as follows: The general wisdom (golden rule) in quantum chaos
is that the quantum diffusion follows the classical diffusion up to the Heisenberg time (or break time, or localization time), defined as
\be \label{tH}
t_H = \frac{2\pi\hbar}{\Delta E},
\ee
where $\Delta E$ is the mean energy level spacing. In our case we have the quasienergies and $\Delta E= \hbar \Delta\omega$, where $\Delta \omega = \Delta \phi/T$, and $\Delta \phi$ is the mean spacing of the eigenphases. This might be estimated at the first sight as $\Delta \phi = 2\pi/N$, but this is an underestimate, as effectively we shall have due to the localization only $\linf$ levels on the interval $[0,2\pi)$. Therefore $\Delta \phi = 2\pi/\linf$ and we find
\be \label{tH2}
t_H = \frac{2\pi T}{\Delta \phi} = T\linf.
\ee
Since $T$ is the period of kicking, and $t_H$ is the real physical continuous time, we get the result that the discrete time (number of iterations of Eq.~(\ref{SM2}) at which the quantum diffusion stops), the localization time $t_{\rm loc}$ is indeed equal to the localization length in momentum space, i.e.
\be \label{tloc=linf}
t_{\rm loc} \approx \linf.
\ee
Since our derivation is not rigorous, we use the approximation symbol rather than equality, in particular as the definition depends linearly on the definition of the Heisenberg time.
Now the final step: By inspection of the dynamics of the Floquet quantal system [Eqs.~(\ref{QKR}),(\ref{Fop})] one can see (see also the derivation in the St\"ockmann's book \cite{Stoe}) that the value of the variance of the angular momentum at the point of stopping the diffusion $\langle (\Delta p)^2\rangle$ is proportional to $\hbar^2 \linf^2$, and to achieve equality we introduce a dimensionless numerical (empirical) factor $\alpha_{\mu}$ by writing $\langle (\Delta p)^2\rangle = \hbar^2 \linf^2 /\alpha_{\mu}$, which on the other hand must be equal just to the classical value at stopping time $t_{\rm loc}$, namely equal to $(I/T)^2D_{\mu} \linf^{\mu}$. From this it follows immediately
\be \label{finallinf}
\linf \approx t_{\rm loc} = \left( \alpha_{\mu} \frac{D_{\mu}(K)}{\tau^2} \right)^{\frac{1}{2-\mu}}.
\ee
The numerical constant $\alpha_{\mu}$ is found empirically by numerical calculations, for instance  in the literature the case $K=5$ with $\mu=1$ is found to be $\alpha_1=0.5$ (however, we find numerically $\alpha_1=0.45$,  taking into account Eq.~(\ref{finallinf}) when studying the model's localization properties). Thus, we have the theoretical formula for the localization length in the case of generalized  classical diffusion [Eqs.~(\ref{varp}),(\ref{varpreal})], which we use in defining the scaling parameter [Eq.~(\ref{MLL:Lamda})].

As for the classical system [Eq.~(\ref{SM2})], we mention that the fraction of the regular part of the classical phase space has been systematically explored using the GALI method \cite{SBA:2007} for the distinction between chaotic and regular classical motion and its quantification for simple (and even for coupled) standard map(s) (see \cite{MSAB2008,MSB2008,MSB2009} and references therein), showing that this fraction decreases with $K$ relatively slowly, then faster around $K_{crit}$, has some smaller oscillations, at $K=5$ still amounts about 2.2\%, and finally for $K\ge 7$ it is zero for all practical purposes (much less than 1\%). Thus, at $K \ge 7 $ we have no problems with the effects of the divided classical phase space. However, there are important subtleties about the classical diffusion process and $D_{\mu} (K)$ which we now discuss.

\begin{figure}
\center
\includegraphics[width=7.5cm]{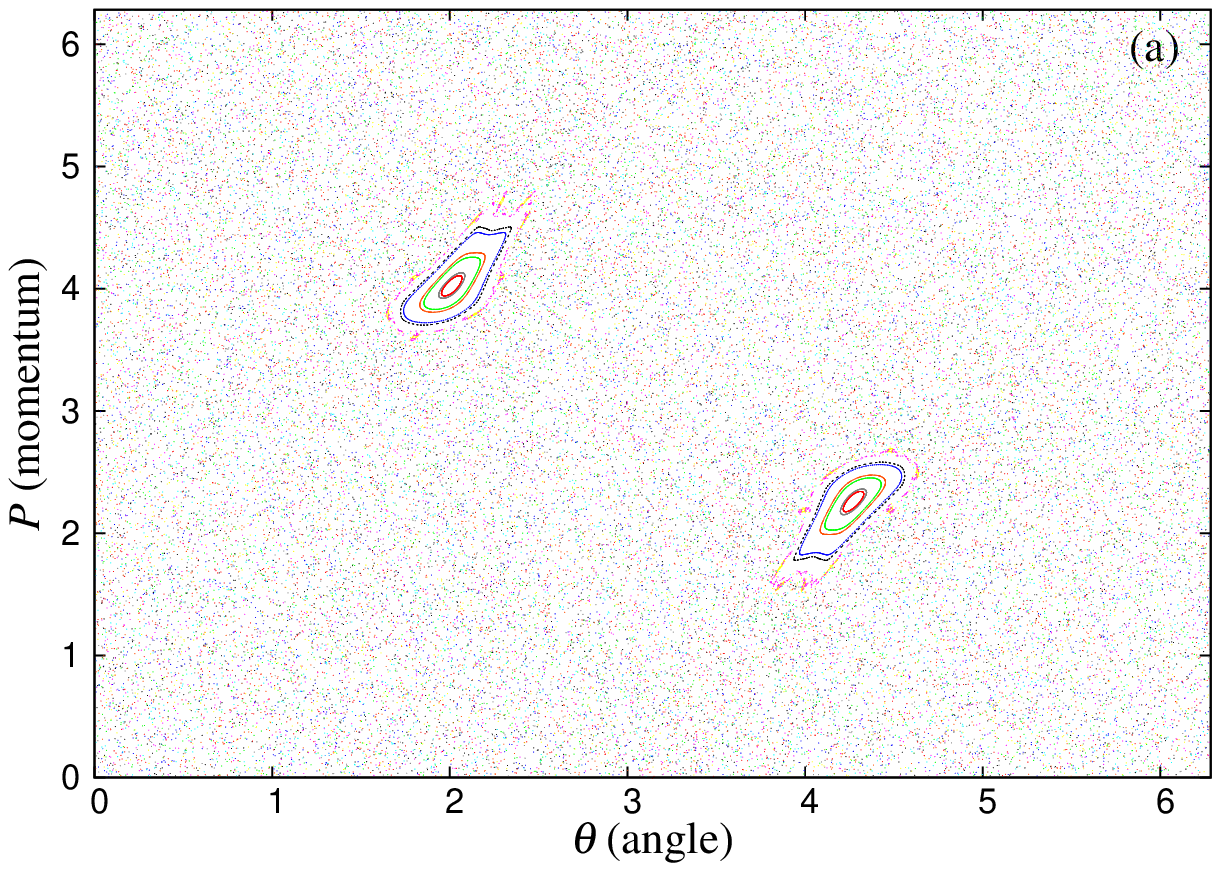}
\includegraphics[width=7.5cm]{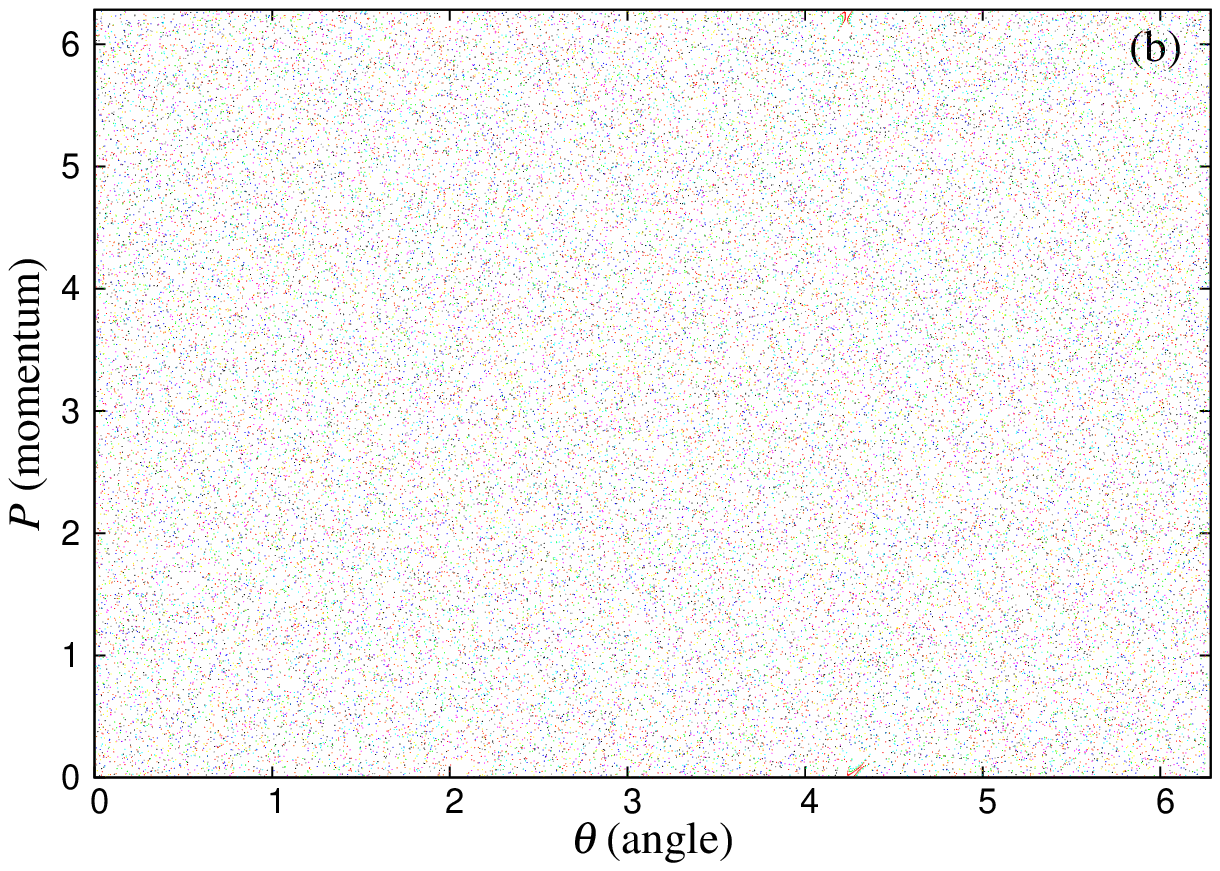}
\caption{[Colour online] The phase portrait of the standard map [Eq.~ (\ref{SM2})] for $K=5$ in (a) and $K=7$ in (b). In (b) we can see two tiny
islands of stability of a period one orbit near $(\theta,P)=(\approx 4.25, 0)$ and $(\theta,P)=(\approx 4.25, 2\pi)$ (see text for more discussion).  The other stable fixed point at $(\theta,P)=(\approx 2.13, 0)$ is hardly visible.}
\label{SMK5}
\end{figure}

We show the phase portraits of the standard map from Eq.~(\ref{SM2}) for $K=5$ and for $K=7$ in Fig.~\ref{SMK5}, in order to demonstrate that at $K=5$ we still have islands of stability of relative area about 2.2\%, which means that the effects of divided phase space cannot be neglected, whilst at $K=7$ there are no large islands of stability. However, there are still two tiny islands of stability around $(\theta,P)=(\approx 4.25, 0)$ and $(\theta,P)=(\approx 4.25, 2\pi)$ (near a period one stable orbit) whose relative fraction in the phase space is found with the GALI method to be $\approx 0.0162\%$. The $K=7$ is the most widely used parameter value for the model in this paper.

In case of the normal diffusion $\mu=1$ the theoretical value of $D_1(K)$ is given in the literature, e.g. in \cite{Izr1990} or \cite{LL1992},
\begin{flalign} \label{Dcl}
D_{1}(K)=
\begin{cases}
 \frac{1}{2} K^2\left [1- 2J_2(K) \left (1-J_2(K) \right ) \right ], \text{if} \ K \ge 4.5 \\
 0.15(K-K_{cr})^3, \text{if} \ K_{cr} < K \le 4.5
\end{cases},
\end{flalign}
where $K_{crit}\simeq 0.9716$ and $J_2(K)$ is the Bessel function. Here we neglect higher terms of order $K^{-2}$. However, there are many important subtle details in the classical diffusion further discussed below.

\begin{figure}
\center
\includegraphics[width=7.5cm]{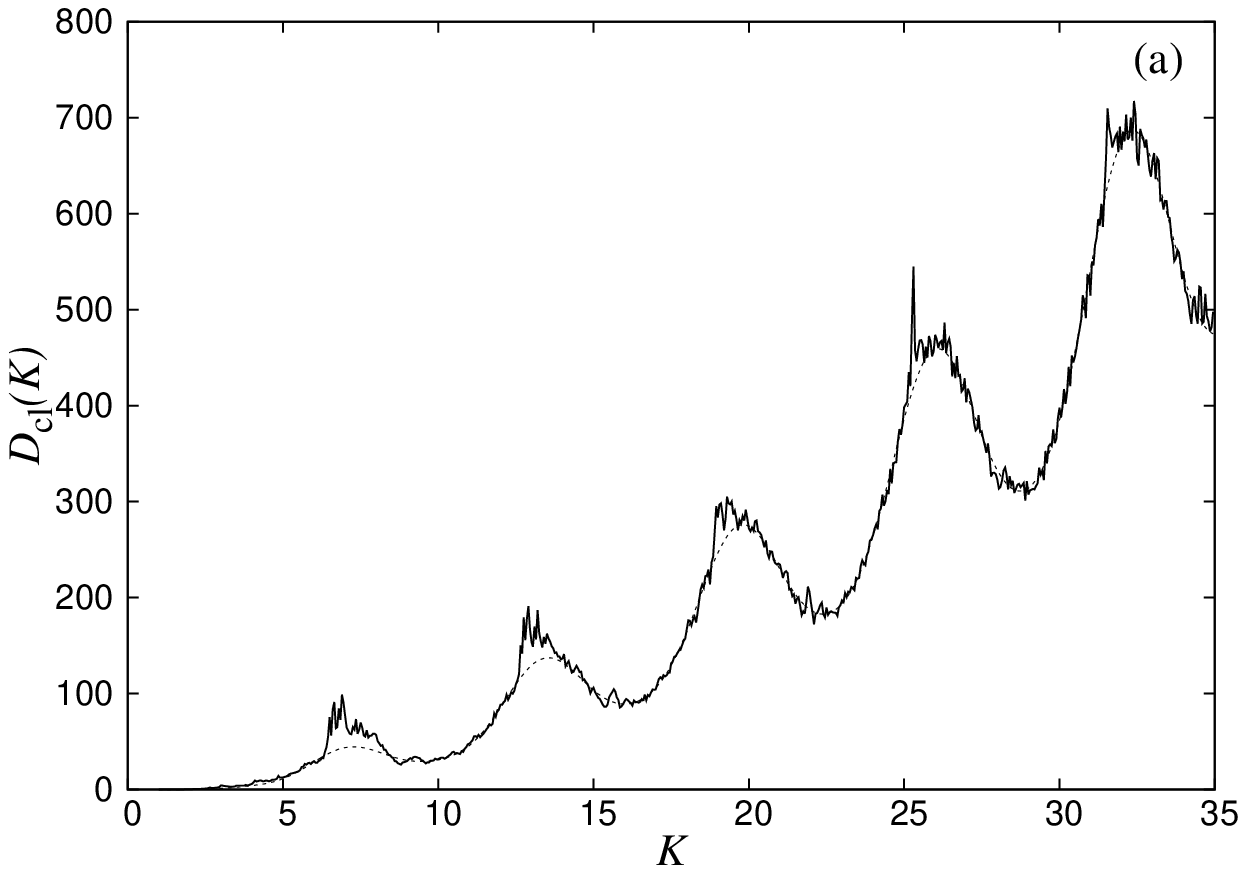}
\includegraphics[width=7.5cm]{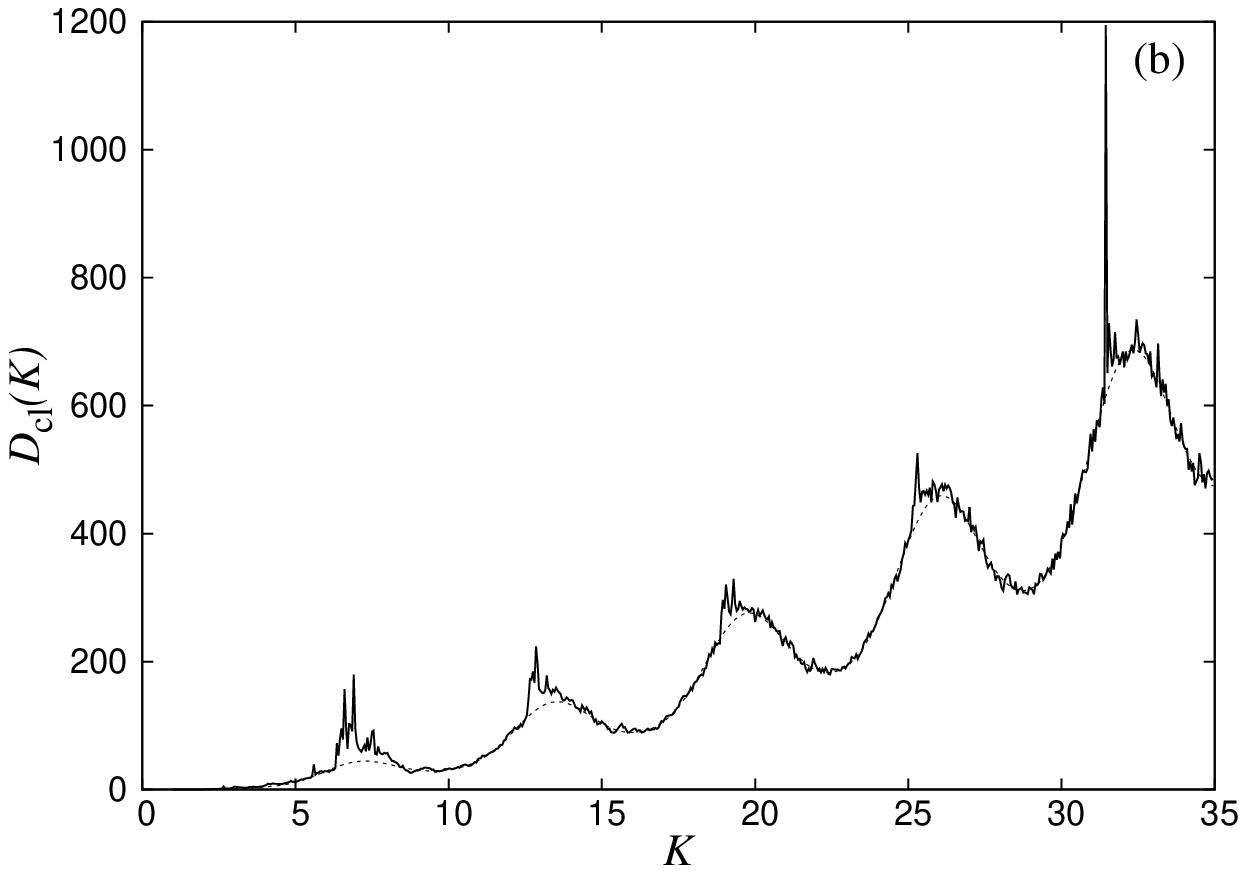}
\caption{Diffusion in the Chirikov map. We show the value of the classical diffusion constant as a function of $K$, for two discrete times $n$, i.e. the number of the iterations of the standard map, $n=1000$ in (a) and $n=5000$ in (b). The smooth background (dotted) agrees perfectly with the theory [Eq.~(\ref{Dcl})], whilst the peaks are due to the anomalous diffusion associated with the accelerator modes and other sticky objects around them.  We have used 5000 initial conditions uniformly distributed in a grid of a square unit $[0,1]\times[0,1]$.}
\label{fig14}
\end{figure}

The dependence of the diffusion constant for the growth of the variance of the momentum on $K$ is very sensitive, and described in the theoretical result [Eq.~(\ref{Dcl})], and fails around the accelerator mode intervals $(2\pi n) \le K \le \sqrt{(2\pi n)^2 +16 }$, $n$ any positive integer. In these intervals for the accelerator modes $n=1$ we have two stable fixed points located at $p=0,\; \theta = \pi -\theta_0$ and $p=0,\; \theta = \pi +\theta_0$, where  $\theta_0 = \arcsin (2\pi/K)$. There are two unstable fixed points at $p=0,\; \theta = \theta_0$ and $p=0,\; \theta = 2\pi - \theta_0$. In our case $K=7$ of Fig.~\ref{SMK5}b we have $\theta_0=1.114$.
 Moreover, as the diffusion might even be anomalous, we have recalculated the diffusion constant numerically, and the results are shown in Fig.~\ref{fig14}. We see that the dotted theoretical curve stemming from Eq.~(\ref{Dcl}) describes the diffusion constant well outside the accelerator mode intervals. In general, however, the diffusion might be non-normal, described in Eq.~(\ref{varp}). For the case $K=7$, which is the main case that we study classically and quantally in this paper, we find three different regimes of diffusion as shown in Fig.~\ref{fig15}. As it will be seen below, for our purposes the middle regime with $\mu\approx 0.9$ and $D_{\mu}\approx 169.82$ is relevant and  important, because $\linf$ is in the range of $N$, the dimension of the $N$-dimensional matrix model introduced in the next section.

\begin{figure}
\center
\includegraphics[width=7.5cm]{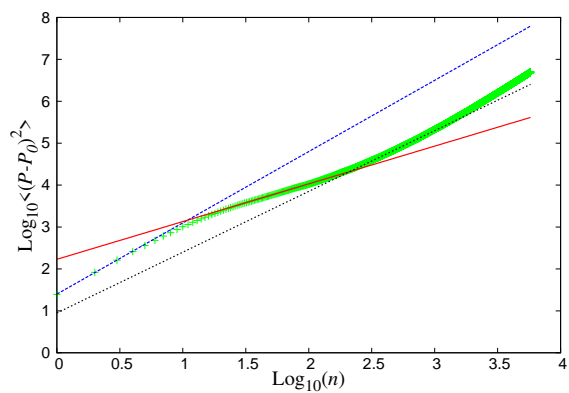}
\caption{The variance of the momentum $P$ in the standard map [Eq.~(\ref{SM2})] with $K=7$ for the same initial conditions as in Fig.~\ref{fig14} as a function of the discrete time $n$ (number of iterations), in log-log representation. The three slopes associated with different types of diffusion are $\mu=1.7$ (dotted), $\mu=0.9$ (solid) and $\mu=1.45$ (dashed). }
\label{fig15}
\end{figure}

\section{The Floquet model system: a finite unitary matrix as
the Floquet operator (Izrailev model) \label{sec2}}

The motion of the QKR [Eq.~(\ref{QKR})] after one period $T$ of the $\psi$ wavefunction can be described also by the following symmetrized Floquet mapping, describing the evolution of the kicked rotator from the middle of a free rotation over a kick to the middle of the next free rotation, as follows
\begin{flalign} \label{Uoper}
& \psi(\theta,t+T) = \hat{U}\psi(\theta,t), \\ \nonumber
& \hat{U} = \exp \left ( i \frac{T\hbar}{4I}\frac{\partial^2}{\partial \theta^2} \right )\exp \left (-i\frac{V_0}{\hbar} \cos \theta \right)\exp \left (i \frac{T\hbar}{4I}\frac{\partial^2}{\partial \theta^2} \right).
\end{flalign}
Thus, the $\psi(\theta,t)$ function is determined in the middle of the rotation, between two successive kicks. The evolution operator $\hat{U}$ of the system corresponds to one period. Due to the instant action of the kick, this evolution can be written as the product of three non-commuting unitary operators, the first and third of which correspond to the free rotation during half a period $\hat{G}(\tau/2)=\exp \left (i\frac{\tau}{4}\frac{\partial^2}{\partial \theta^2} \right)$, $ \tau \equiv \hbar T/I$, while the second $\hat{B}(k)=\exp(-ik\cos \theta)$, $ k \equiv V_0/\hbar$, describes the kick. The system's behavior depends only on two dimensionless parameters, namely $\tau$ and $k$, and its correspondence with the classical system is described by the relation $K=k \tau=V_0 T/I$. In the case $K\equiv k \tau \gg 1$ the motion is well known to be strongly chaotic,
for $K\ge 7$ almost without any regular islands of stability, as explained in the previous section. The transition to classical mechanics is described by the limit $k \rightarrow \infty$, $\tau \rightarrow 0$ while $K=\rm{const}$. We shall consider mostly the semiclassical regime $k\ge K$, where $\tau \le 1$.

In order to study how the localization affects the statistical properties of the quasienergy spectra, we use the model's representation in the momentum space with a finite number $N$ of levels \cite{Izr1988,Izr1990,Izr1986,Izr1987,Izr1989}
\begin{flalign} \label{u_repres}
u_n(t+T) = \sum_{m=1}^{N} U_{nm}u_m(t), \ n,m=1,2,...,N \enspace .
\end{flalign}
The finite symmetric unitary matrix $U_{nm}$ determines the evolution of an $N$-dimensional vector, namely the Fourier transform $u_n(t)$ of $\psi(\theta,t)$, and is composed in the following way
\be \label{Unm}
    U_{nm}=\sum_{n'm'}G_{nm'}B_{n'm'}G_{n'm},
\ee
where $G_{ll'}=\exp \left (i\tau l^2/4 \right )\delta_{ll'}$ is a diagonal matrix corresponding to free rotation during a half period $T/2$, and  the matrix $B_{n'm'}$ describing the one kick has the following form
\begin{flalign} \label{Bnmoper}\nonumber
  & B_{n'm'}= \frac{1}{2N+1}\times \\ \nonumber
  & \sum_{l=1}^{2N+1} \left \{ \cos \left [ \left (n'-m' \right ) \frac{2 \pi l}{2N+1}\right ] - \cos \left [(n'+m')\frac{2 \pi l}{2N+1} \right ] \right \} \\
  & \times  \exp \left [-ik\cos \left (\frac{2 \pi l}{2N+1}\right ) \right ].
\end{flalign}
The model in Eqs.~(\ref{u_repres}-\ref{Bnmoper}), which we refer to as
Izrailev model, with a finite number of states is considered as the quantum analogue of the classical standard mapping on the torus with closed momentum $p$ and phase $\theta$, where $U_{nm}$ describes only the odd states of the systems, i.e. $\psi(\theta)=-\psi(-\theta)$, provided we have the case of the quantum resonance, namely $\tau =4\pi r/(2N+1)$, where $r$ is a positive integer, as in Eq.~(\ref{Restau}). The matrix [Eq.~(\ref{Bnmoper})] is obtained by starting the derivation from the odd-parity basis of $\sin(n\theta)$ rather than the general angular momentum basis $\exp(in\theta)$.

Nevertheless, we shall use this model for any value of $\tau$ and $k$, as a model which in the resonant and in the generic case (irrational $\tau/(4\pi)$) corresponds to the classical kicked rotator, and in the limit $N\rightarrow \infty$ approaches the infinite dimensional model [Eq.~(\ref{Uoper})], restricted to the symmetry class of the odd eigenfunctions. It is of course just one of the possible discrete approximations to the continuous infinite dimensional model.

\begin{figure}
\center
\includegraphics[width=7.5cm]{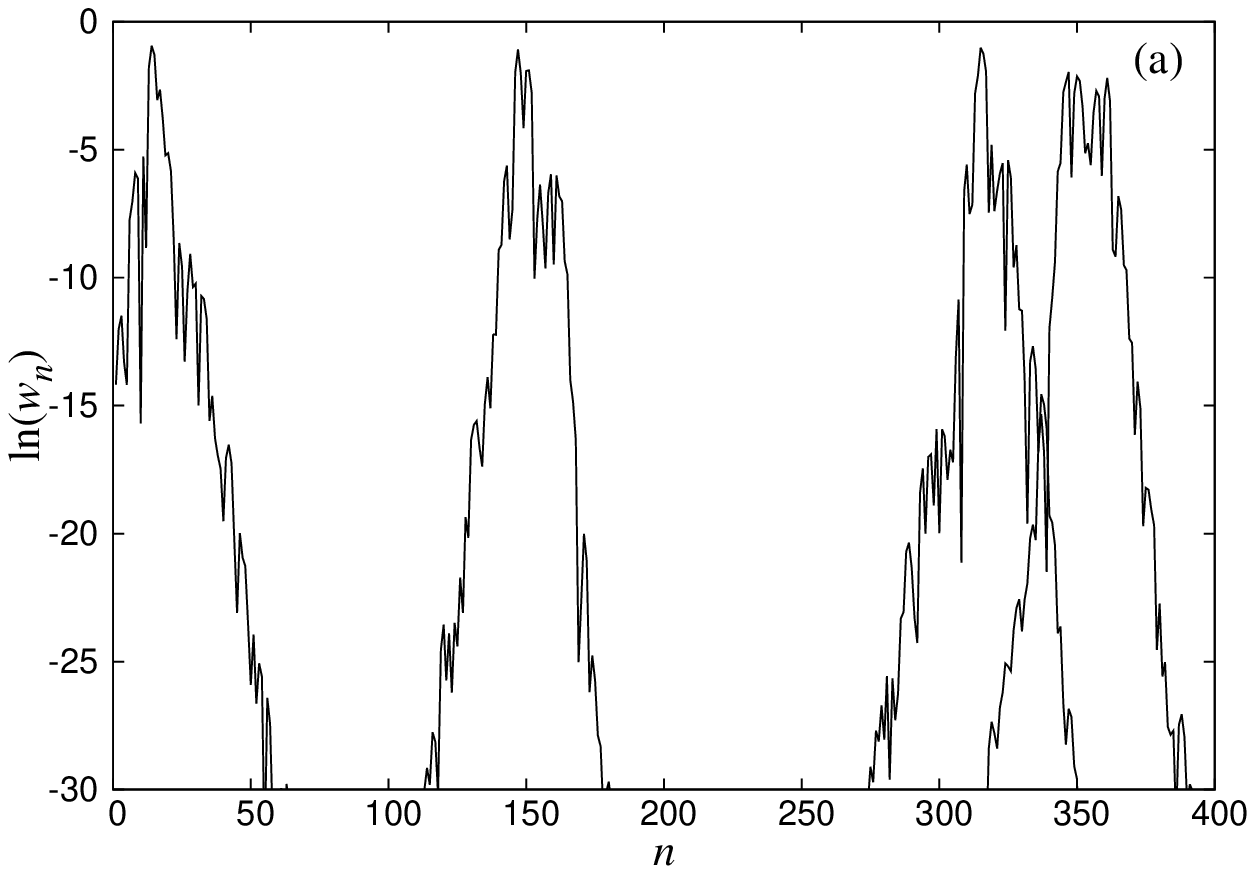}
\includegraphics[width=7.5cm]{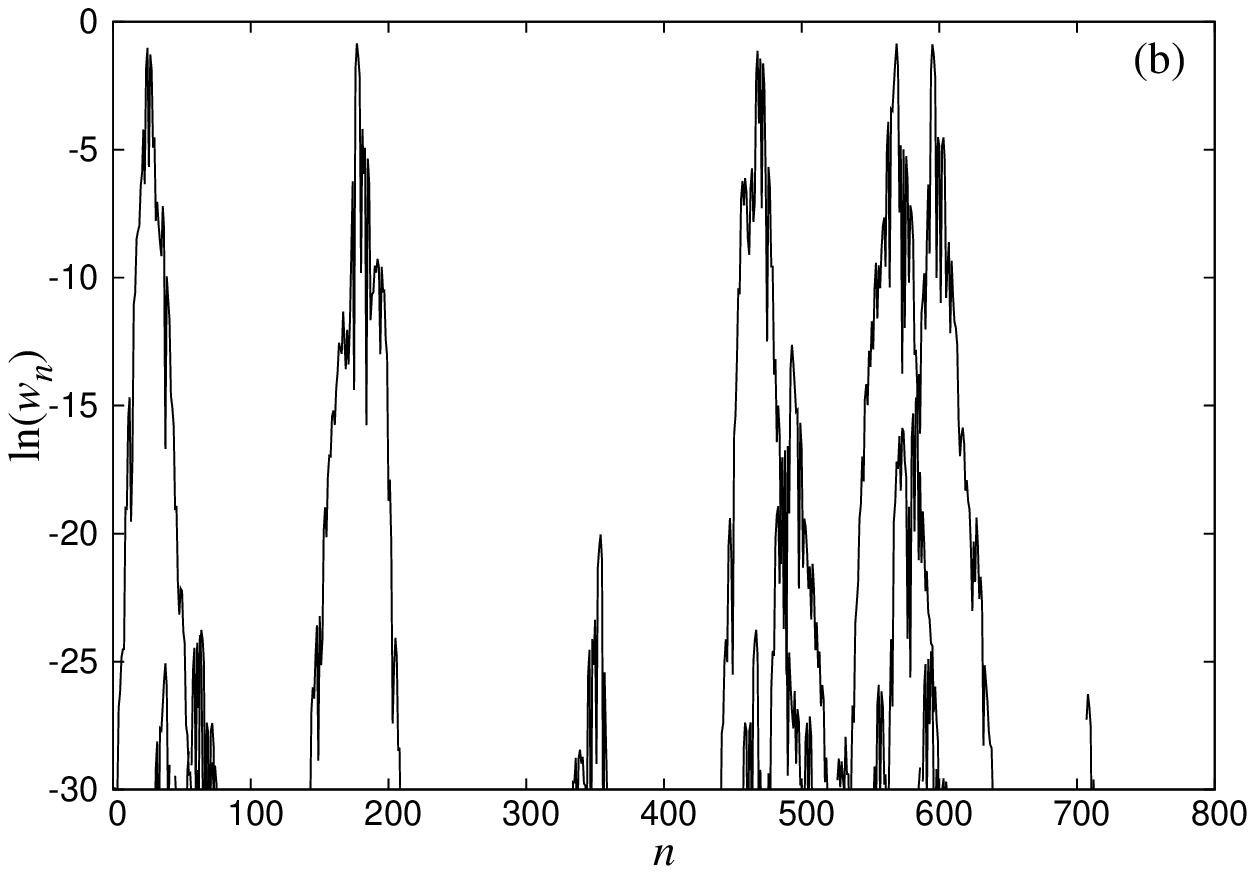}
\caption{(a) A sample of strong localized eigenstates for $K=7$, $r=222$, $k\approx 2.00$ and $N=398$ (b) Same for $K=7$, $r=444$, $k\approx 2.00$ and $N=796$.}
\label{ppfig1}
\end{figure}

The difference of behaviour between the generic case and the quantum resonance shows up only at very large times, which grow fast with $(2N+1)$, as explained in section \ref{secmod}.  It turns out that also the eigenfunctions and the spectra of the eigenphases at finite dimension $N$ of the matrices that we consider do not show any significant differences in structural behaviour for the rational or irrational $\tau/(4\pi)$, which we have carefully checked. Indeed, although the eigenfunctions and the spectrum of the eigenphases exhibit {\em sensitive dependence on the parameters} $\tau$ and $k$, their statistical properties are stable against the small changes of $\tau$ and $k$. This is an advantage, as instead of using very large single matrices for the statistical analysis, we can take a large ensemble of smaller matrices for values of $\tau$ and $k$ around some central value of $\tau=\tau_0$ and $k=k_0$, which greatly facilitates the numerical calculations  and improves the statistical significance of our empirical results. Therefore our approach is physically meaningful. Similar approach was undertaken by Izrailev (see \cite{Izr1990}
and references therein). In Fig.~\ref{ppfig1} we show the examples of strongly exponentially localized eigenstates by plotting the natural logarithm of the
probabilities  $w_n=|u_n|^2$ versus the momentum quantum number $n$, for two different matrix dimensions $N$. By calculating the localization length $l_{\infty}$ from the slopes $\sigma$ of these eigenfunctions using Eq.~(\ref{exploc}) we can get the first quantitative empirical localization
measure to be discussed and used later on. Here $l_{\infty}=2/\sigma \approx 2.5  \ll N=398$ for Fig.~\ref{ppfig1}a and $\approx 2.2 \ll N=796$ for Fig.~\ref{ppfig1}b.

\begin{figure}
\center
\includegraphics[width=7.5cm]{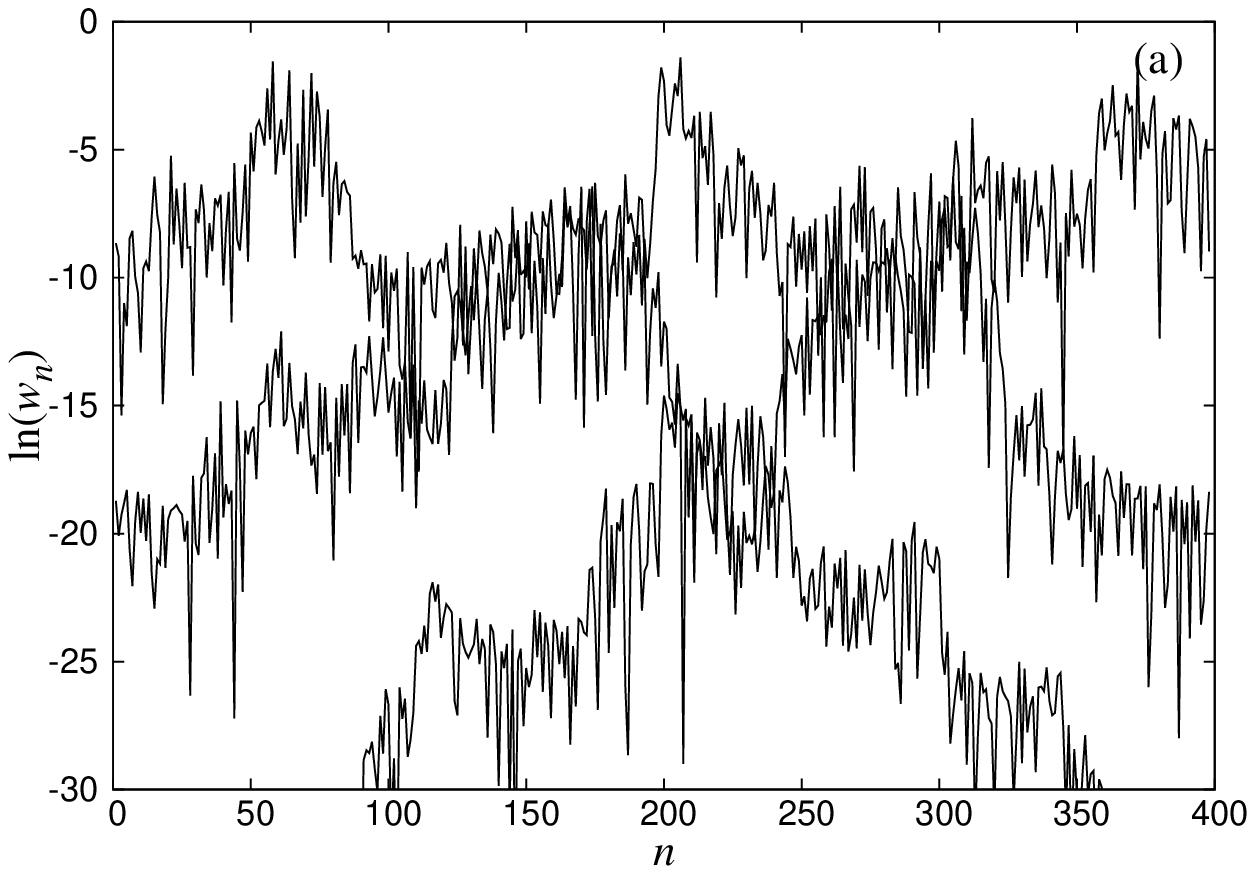}
\includegraphics[width=7.5cm]{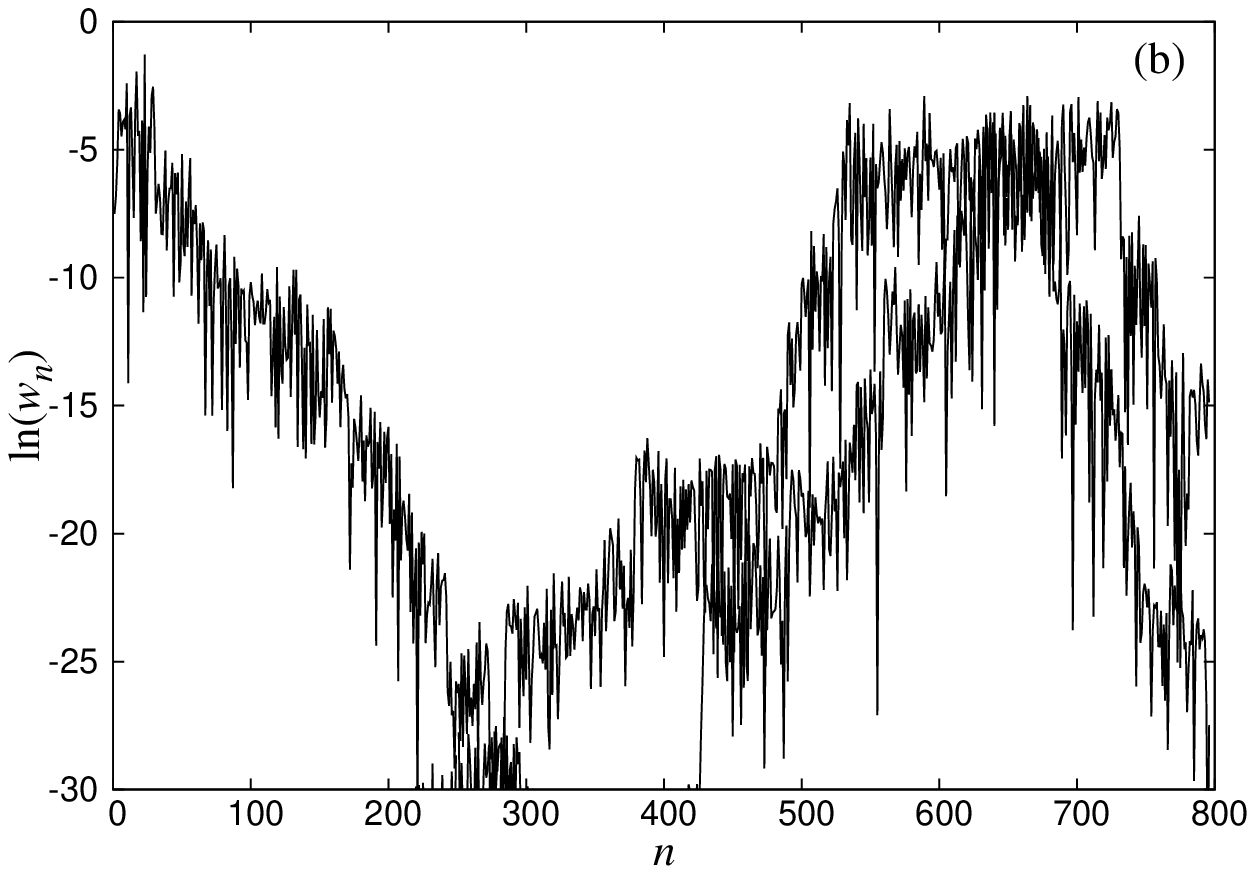}
\caption{(a) A sample of weak localized eigenstates for $K=7$, $r=63$, $k\approx 7.05$ and $N=398$ (b) Same for $K=7$, $r=127$, $k\approx 6.99$ and $N=796$.}
\label{fig1}
\end{figure}

At larger $k$ the localization length can become comparable to $N$ and the size effects start to play a role, therefore $l_{\infty}$ is more difficult to determine as the fluctuations of $2/\sigma$ are larger, but still can be done to some extent. Such a case is shown in Fig.~\ref{fig1}a with $l_{\infty}=2/\sigma \approx 57 \ll N=398$ and in Fig.~\ref{fig1}b
$\approx 20 \ll N=796$, where the quantification is difficult even if the localization length $2/\sigma$ is well below $N$. We must be aware of the fact that the fluctuations in $2/\sigma$ are very large, as was observed already in the pioneering works of Chirikov, Casati, Izrailev and Shepelyansky.

Nevertheless, as long as the localization length is small enough, it is correctly predicted by the theory [Eq.~(\ref{finallinf})], and is independent of the dimension $N$. As $N$ increases with $\linf$ being fixed we approach the regime of strong localization and Poissonian statistics for the eigenvalues. Only when $\linf$ becomes comparable to $N$ or larger, we observe gradual transition to the full quantum chaoticity, namely to GOE (or COE) behaviour, and anything in between. This transition is of central interest and is the main subject of the next sections.

\begin{figure}
\center
\includegraphics[width=7.5cm]{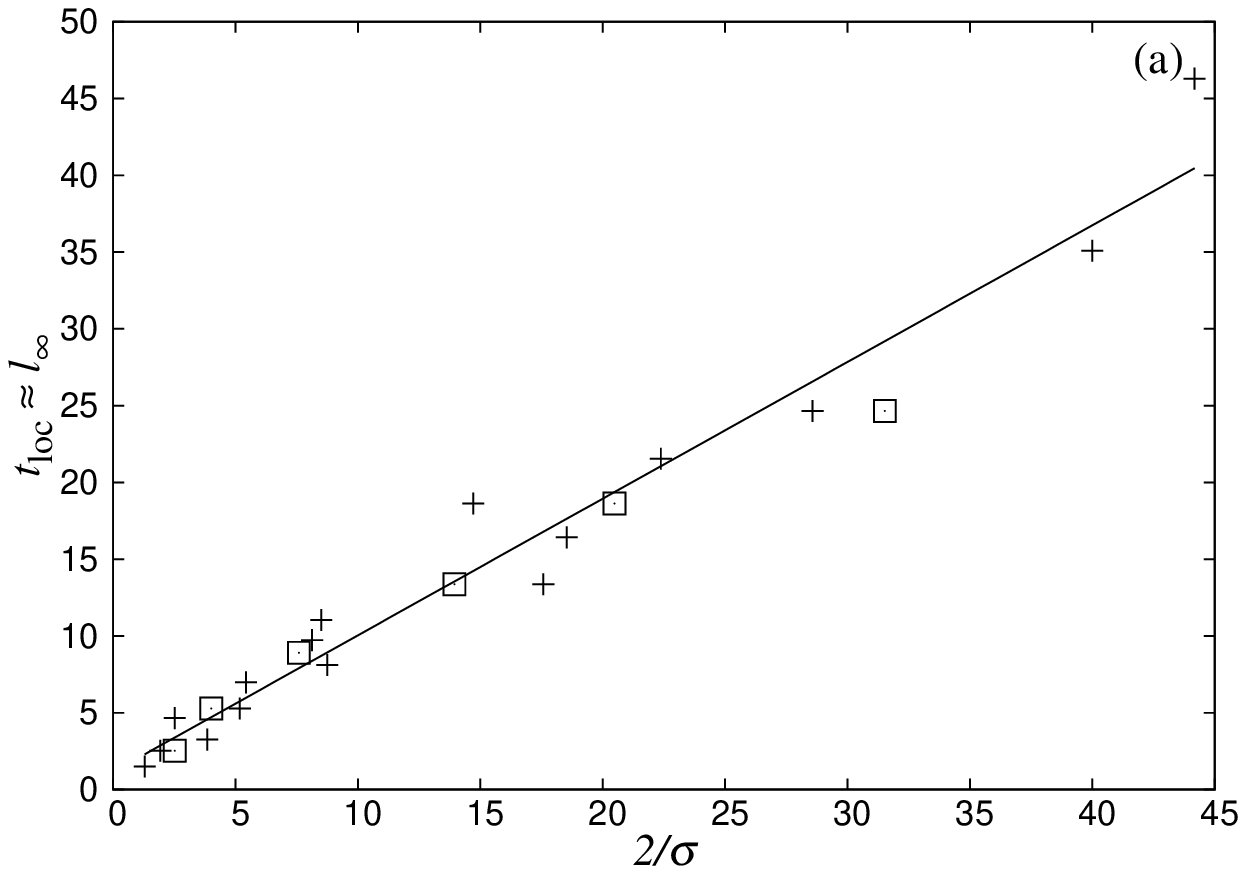}
\includegraphics[width=7.5cm]{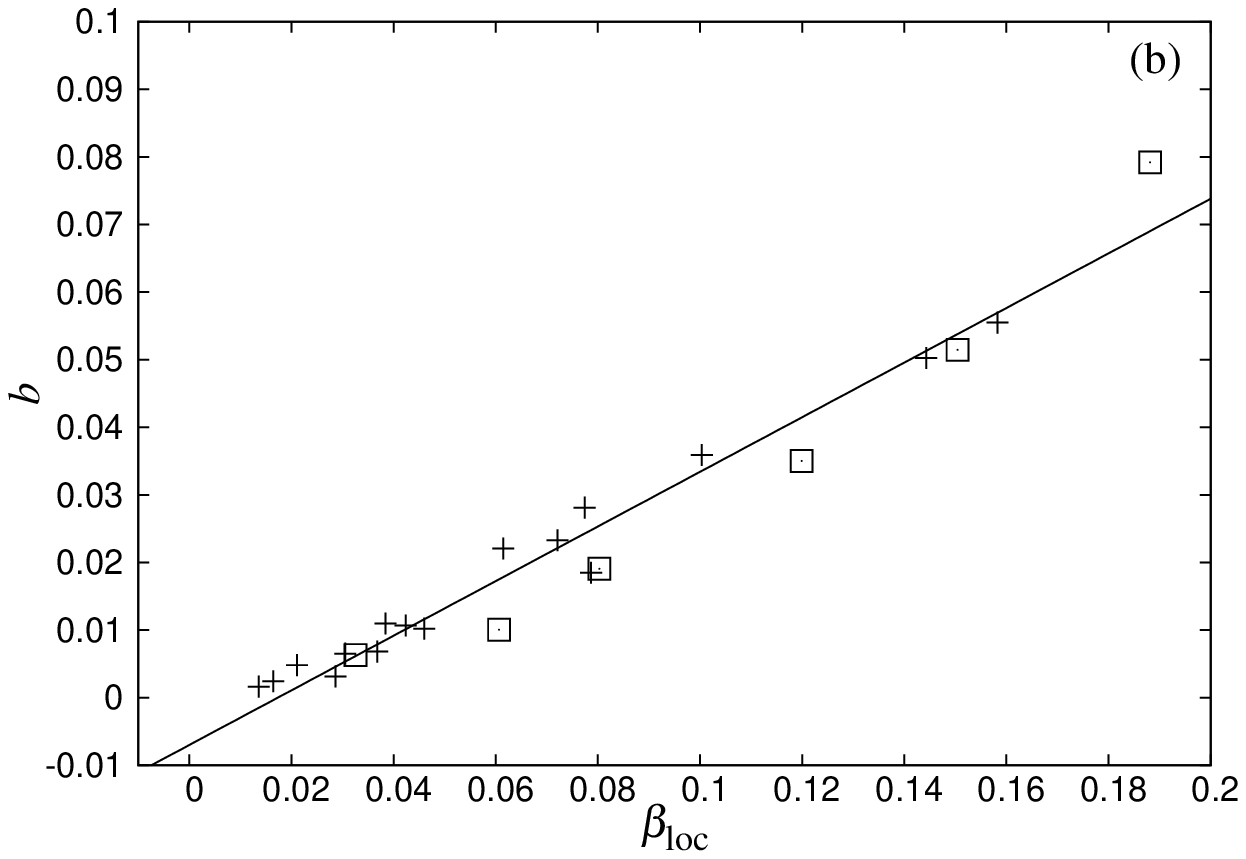}
\caption{$2/\sigma$ vs $\l_{\infty}$,
where $\sigma$ is the slope of the eigenfunctions in momentum space, namely
of $w_n=|u_n|^2$ vs n, for several $k$-values and $N=398(\square)$ and $N=796
(+)$ (fit slope $\approx 0.89$).
(b) $\bex$ vs. $\blo$, $N=398(\square)$ and $N=796(+)$
(fit slope $\approx 0.404$).}
\label{fig2}
\end{figure}

In Fig.~\ref{fig2}a we show the relationship between the $2/\sigma$ and the theoretical $l_{\infty}$ for variety of matrices [Eqs.~(\ref{u_repres}-\ref{Bnmoper})] and $K=7$. We have statistically significant linear relationship with the slope $\approx 0.9$ close to unity (identity). Thus the theoretical $\linf$ from Eq.~(\ref{finallinf}) agrees reasonably with the empirical localization length $2/\sigma$ as long as they are both sufficiently smaller than $N$.

We can define also the relative exponential localization measure, defined as the ratio of $2/\sigma$ and $N$,
\be \label{bex}
\bex = \frac{2}{N\sigma},
\ee
which has meaning only if $\bex \ll 1$, because $2/\sigma$ is well defined only if it is much smaller than $N$, but even then we must be aware of large statistical fluctuations  of $2/\sigma$.

There is another way to empirically quantify the degree of localization based on the information entropy, less sensitive to the finite size effects, denoted by $\blo$, and is discussed in the next section. In Fig.~\ref{fig2}b we show the relationship between $\bex$ and $\blo$. It is seen to be a linear relationship, but it must be emphasized that it applies only to sufficiently small values of $\bex$, because empirical exponential localization length $2/\sigma$ completely loses its meaning at values $\ge N$, or in fact, even much earlier. This is the reason why the slope in Fig.~\ref{fig2}b $\sigma\approx 0.404$ does not agree with Eq.~(\ref{bla}) of section \ref{sec5}, namely the slope there is $1/\gamma\approx 0.25$, $\gamma=4.04$, which is due to large fluctuations in $\sigma$.

\section{Dynamical (Chirikov) localization of the eigenstates and its measure \label{sec3}}

Based on the examples of the eigenstates shown in the previous section, and following \cite{Izr1990} and the references therein, we introduce the information entropy of the eigenstates as follows.

For each $N$-dimensional eigenvector of the matrix $U_{nm}$ the information entropy is
\be  \label{infoentr}
 \mathscr{H}_N(u_1,...,u_N) = -\sum_{n=1}^{N}w_n \ln w_n,
\ee
where $w_n = |u_n|^2$, and $\sum_n |u_n|^2 = 1$.

In case of the random matrix theory being applicable to our system [Eqs.~(\ref{Uoper}) and (\ref{u_repres}-\ref{Bnmoper})], namely the COE (or GOE), due to the isotropic distribution of the eigenvectors of a COE of random matrices,  we have the probability density function of $|u_n|$ on the interval $[0,1]$,
\begin{flalign} \label{GOEeigvec}
w_N(|u_n|) = \frac{2\Gamma(N/2)}{\sqrt{\pi} \Gamma((N-1)/2) } (1-|u_n|^2)^{(N-3)/2}.
\end{flalign}
It is easy to show that in the limit $N\rightarrow \infty$ this becomes a Gaussian distribution
\be \label{Gaussianeigvec}
w_N(|u_n|) = \sqrt{\frac{2N}{\pi}} \exp \left(-\frac{N|u_n|^2}{2}
\right),
\ee
and the corresponding information entropy [Eq.~(\ref{infoentr})] is equal to
\begin{flalign} \label{H_GOE}
 \mathscr{H}_{N}^{GOE}=\psi \left (\frac{1}{2}N+1 \right )-\psi \left (\frac{3}{2} \right )\simeq \ln \left (\frac{1}{2}Na \right )+O(1/N),
\end{flalign}
where $a=\frac{4}{\exp(2-\gamma)}\approx 0.96$, while $\psi$ is the digamma function and $\gamma$ the Euler constant ($\simeq 0.57721...$). For a uniform distribution over $M$ states $w_n=1/M$ we get $\mathscr{H}_{N} \approx \log M$, and thus $M \approx \exp (\mathscr{H}_{N})$. Thus, we get the insight that the correct measure of localization must be proportional to $\exp (\mathscr{H}_{N})$, but properly normalized, such that in case of extendedness (GOE/COE) it is equal to $N$.

Therefore the {\it entropy localization length} $l_H$ is defined as \be\label{lh:eq}
 l_H=N \exp \left (\mathscr{H}_{N}-\mathscr{H}_{N}^{GOE} \right ).
\ee
Indeed, for entirely extended eigenstates $l_H=N$. Thus, $l_H$ can be calculated for every eigenstate individually. However, all eigenstates,
while being quite different in detail, are exponentially localized,
and thus statistically very similar. Therefore, in order to minimize the fluctuations one uses the {\it mean localization length} $d\equiv \langle l_H \rangle$, which is computed by averaging the entropy over all eigenvectors of the same matrix (or even over an ensemble of similar matrices)
\be\label{d:eq}
 d \equiv \langle l_H \rangle = N \exp \left (\langle \mathscr{H}_{N} \rangle-\mathscr{H}_{N}^{GOE} \right ).
\ee
The {\it localization parameter} $\blo$ is then defined as
\be\label{beta_loc:eq}
    \blo=\frac{d}{N}\equiv \frac{\langle l_H\rangle}{N}.
\ee
Its relationship to $\bex$ at small values $\blo$ is shown in Fig.~\ref{fig2}b and discussed in the previous section.

The parameter that determines the transition from weak to strong quantum chaos is neither the strength parameter $k$ nor the localization length $l_{\infty}$, but the ratio of the localization length $l_{\infty}$ to the size $N$ of the system in momentum $p$
\be\label{MLL:Lamda}
  \La=\frac{l_{\infty}}{N} = \frac{1}{N}
\left( \frac{\alpha_{\mu} D_{\mu} (K)}{\tau^2}\right)^{\frac{1}{2-\mu}} ,
\ee
where $l_{\infty} \approx t_{\rm loc}$, the theoretical localization length, was derived in Eq.~(\ref{finallinf}). $\La$ is the scaling parameter of the system. This is one of the main results of the present work as it incorporates
normal diffusion $\mu=1$ and the anomalous diffusion $\mu\not=1$. The relationship of $\La$ to $\blo$ is discussed in section \ref{sec5}.

\section{The quasienergy spectrum and its statistical properties \label{sec4}}

In this section we study the statistical properties of the spectrum of the eigenvalues $\lambda_j$ corresponding to the eigenstates $u_n^{(j)}$, labeled by $j$, of the Floquet operator $F_{nm}$ introduced in Eq.~(\ref{Fmatrix}), namely
\be \label{Feigen}
\lambda_j u_n^j = \sum_{-\infty}^{\infty} F_{nm} u_m^j,
\ee
where due to the unitarity of $F_{nm}$ all eigenvalues $\lambda_j$ must be on
the complex unit circle, $\lambda_j=\exp (i\phi_j)$. Thus, our analysis is concerned about the statistical properties of the spectrum of the eigenphases $\phi_j \in [0,2\pi)$. This is equivalent to the quasienergies $\hbar \omega_j= \hbar\phi_j/T$ \cite{Zel1966}, where $T$ is the period of the time-periodic Hamiltonian. As mentioned in section \ref{secmod}, the spectrum can be continuous (in case of the quantum resonance with rational $\tau/(4\pi)= r/q$, having extended states, and no dynamical localization), or discrete (in the generic case of sufficiently irrational $\tau/(4\pi)$ ) with dynamical localization. Even then, the spectrum has infinite level (eigenvalue) density as we have infinitely many eigenphases $\phi_j$ on the interval $[0,2\pi)$, and consequently all level spacings are zero. However, for any finite dimension $N$, no matter how large, the mean level spacing is $2\pi/N$, and thus finite, and we can begin the statistical analysis of the quasienergy spectrum.

Therefore, instead of using the infinite dimensional system [Eq.~(\ref{Fmatrix})], we study the finite dimensional Izrailev model [Eqs.~(\ref{u_repres}-\ref{Bnmoper})]. It is a discrete approximation to the exact initial infinite dimensional system [Eq.~(\ref{Fop})], or in its symmetrized version [Eq.~(\ref{Uoper})]. In this case we can best use irrational $\tau/(4\pi)$, to ensure that we are away from the quantum resonance.

The most important statistical measure of the eigenvalues is the level spacing distribution. We study only the level spacing distribution, but in three different ways of analyzing it, namely as the probability density function $P(S)$, its cumulative distribution function $W(S)$ and the so-called $U$-function, introduced by Prosen and Robnik in \cite{ProRob1994a,ProRob1994b} (see Appendix \ref{ap:A}).

In so doing we come to the central point of this work, namely the empirical
almost identity relationship between localization measure $\blo$ of the eigenstates and the spectral level repulsion parameter (exponent) in the level spacing distribution $P(S)$, proposed already by Izrailev in \cite{Izr1988}, \cite{Izr1989} and \cite{Izr1990}. The localization length $\linf$, derived by the semiclassical argument [Eq.~(\ref{finallinf})], through the $\blo$ - $\La$ relationship directly gives a prediction for the level repulsion parameter in the semiclassical regime of $k \ge K$ and large $N$.

Here we reproduce all Izrailev's  findings (for $K=5$), generalize them (for many other values of $K$, predominantly at $K=7$), sharpen his results and put them in the broader perspective including the autonomous (time-independent) Hamiltonian systems. We find that the Brody distribution captures the numerically found level spacing distribution statistically highly significantly, notably better than the distribution function proposed in \cite{Izr1988,Izr1989,Izr1990}, also \cite{CIM1991}, and this is entirely in line with the results on the dynamically localized chaotic eigenstates in time-independent Hamiltonian systems, like billiards \cite{BatRob2010,BatRob2012,BatManRob2013}, even in the mixed type regime (after separating the regular and chaotic eigenstates).

\subsection{Level spacing distribution: $P(S)$, $W(S)$ and $U(S)$ \label{subsec4.1}}

To study the eigenvalue statistics of quantum Floquet systems and quantum maps one considers the eigenphases $\phi_n \in [0,2\pi)$ defined by $\lambda_n=e^{i\phi_n}$. In such case the spectral unfolding procedure is very easy, as the mean level density is $N/(2\pi)$, i.e. the mean level spacing is $2\pi/N$. The histogram of the level spacing distribution $P(S)$ is the
distribution of the spacings $S_n:=\frac{N}{2\pi}(\phi_{n+1}-\phi_n)$ with $n=1,...,N$ and $\phi_{N+1}:=\phi_1$, in the bins of certain suitable size $\Delta S$. The factor $N/2\pi$ ensures that the average of all spacings $S_n$ is 1, and thus $P(S)$ is supported on the interval $[0,N]$, and its upper limit goes to $\infty$ when $N\rightarrow \infty$.

The cumulative distribution $W(S)$, or integrated level spacing distribution, preserving the full accuracy of all numerical eigenvalues/spacings, useful especially when the number of levels $N$ is small, is defined as
\be\label{CDF}
    W(S)=\int_0^SP(x)dx \equiv \frac{\# \{n | S_n \leq S\}}{N}.
\ee
Finally, we shall use also the so-called $U$-function (see the Appendix \ref{ap:A})
\be\label{Ufun}
    U(S)=\frac{2}{\pi} \arccos \sqrt{1-W(S)}.
\ee
The $U$-function has the advantage that its expected statistical error $\delta U$ is independent of $S$, being constant for each $S$ and equal to $\delta U=1/(\pi \sqrt N_s)$, where $N_s$ is the total number of objects in the $W(S)$ distribution. The numerical pre-factor $2/\pi$ in Eq.~(\ref{Ufun}) is determined in such a way that $U(S)$ $\in [0,1]$ when $W(S)$ $\in [0,1]$. The $U$-function is an excellent and refined criterion used to assess the goodness of the fit of the theoretical level spacing distribution.

The important special level spacing distributions that we are using are the
following.
\begin{itemize}
  \item The Poisson distribution
  \be\label{PoiDistrib}
    P_{\rm Poisson}(S)= e^{-S}, \quad W_{\rm Poisson}(S)=1- e^{-S}.
  \ee
  \item The COE or GOE distribution
  \bea\label{COEDistrib} \nonumber
    P_{\rm COE}(S) & \approx & P_{\rm Wigner}(S)= \frac{\pi}{2}\exp \left (-\frac{\pi}{4}S^2 \right ),\\
    W_{\rm COE}(S) & = & 1-\exp \left (\frac{-\pi S^2}{4} \right ).
  \eea
  \item The Brody distribution \cite{Bro1973,Bro1981}
  \be \label{BrodyDistrib}
    P_{\rm BR}(S)=C_1 S^\beta \exp \left (-C_2 S^{\beta+1} \right ),
  \ee
 where the two parameters $C_1$ and $C_2$ are determined by the two generic normalization conditions that must be obeyed by any $P(S)$,
 \be\label{NormCond}
    \int_0^\infty P(S) dS=1, \quad \langle S \rangle=\int_0^\infty S P(S) dS=1,
 \ee
 thus with $\langle S \rangle=1$ being the mean distance between neighboring levels (after unfolding). Hence
 \be\label{C1C2}
    C_1=(\beta+1)C_2, \quad C_2=\left[\Gamma\left(\frac{\beta+2}{\beta+1}\right)\right]^{\beta+1},
 \ee
 where $\Gamma(x)$ denotes the Gamma function. In the strongly localized regime $\beta=0$ we observe Poissonian statistics while in the fully chaotic one $\beta=1$ and the RMT applies. The Brody cumulative level spacing distribution is
 \be \label{BrodyCum}
    W_{\rm BR}(S)=1-\exp(-C_2 S^{\beta+1}).
 \ee
  \item Izrailev distribution: In \cite{Izr1988,Izr1989}, Izrailev suggested the following distribution in order to describe the intermediate statistics, i.e. the non-integer $\beta$ in the following PDF could be associated with the statistics of the quasienergy states with chaotic localized eigenfunctions, also approximating the level spacing distribution arising from the Dyson
      COE joint probability distribution \cite{Por1965}
      \begin{flalign}\label{IzrailevDistrib}
      P_{\rm IZ}(S) = A\left (\frac{1}{2}\pi S \right )^{\beta} \exp \left [-\frac{1}{16}\beta \pi^2 S^2 -\left (B-\frac{1}{4}\pi \beta \right )S \right],
      \end{flalign}
      where again the two parameters $A$ and $B$ are determined by the two normalization conditions $\langle 1 \rangle = \langle S \rangle = 1$ given above.
\end{itemize}

Of course, we must be fully aware of the fact that both, Brody and Izrailev distributions, are approximations. It is clear that at $\bro=1$ we get precisely Wigner surmise [Eq.~(\ref{COEDistrib})], which is the exact GOE only for two-dimensional Gaussian random matrices, and thus only an (excellent) approximation for the infinite dimensional GOE case. Indeed, if we try to fit the exact infinite dimensional GOE level spacing distribution with the Brody distribution, we do not get $\bro=1$, but instead $\bro=0.953$, see \cite{Bro1981}. Also, we should mention that Izrailev et al have published an improved distribution function \cite{CIM1991}, which we have also tested, and is defined by
\begin{flalign} \label{NewIzr}
P_{\rm IZ}^{\rm new}(S) =
A S^{\beta}(1+B \beta S)^{f(\beta)}\exp[ -\frac{\pi^2}{16} \beta S^2 - \frac{\pi}{2}(1-\frac{\beta}{2}S)]
\end{flalign}
where $f(\beta)=\frac{2^{\beta}(1-\frac{\beta}{2})}{\beta}-0.16874$
and $A, B$ are the normalization parameters. We found (see below) that in our applications it is even worse than the original version (\ref{IzrailevDistrib}).

\subsection{Analysis of the level spacing distribution of numerical spectra \label{subsec4.2}}

For the numerical calculations and results regarding the spacing distributions $P(S)$ (and $W(S)$) for the eigenphases $\phi_j$, we have considered a range of $41$-values of the quantum parameter $k$ ($=2,3,...,42$) keeping fixed the classical parameter $K=7$ (where the phase space is fully chaotic as shown in section \ref{secmod}). In order to ameliorate the statistics we considered a sample of 161 matrices $U_{nm}$ of size $N=398$ ($\approx 64,000$ elements), in a similar manner as in \cite{Izr1988}) with slightly different values of $k$ (with the step size $\Delta k = \pm 0.00125 \ll k$). For some samples we reached up to 641 matrices $U_{nm}$ of size $N=398$ acquiring qualitatively the same results.

For the ensemble of $M=641$ matrices of size $N=398$, in case $K=7$ and $k=11$,  using the $\chi^2$ best fitting procedure (described in more detail below) we found $\bro=0.421$ and all three representations clearly show excellent agreement with the best fitting Brody distribution.
\begin{figure*}
\center
\includegraphics[width=7.5cm]{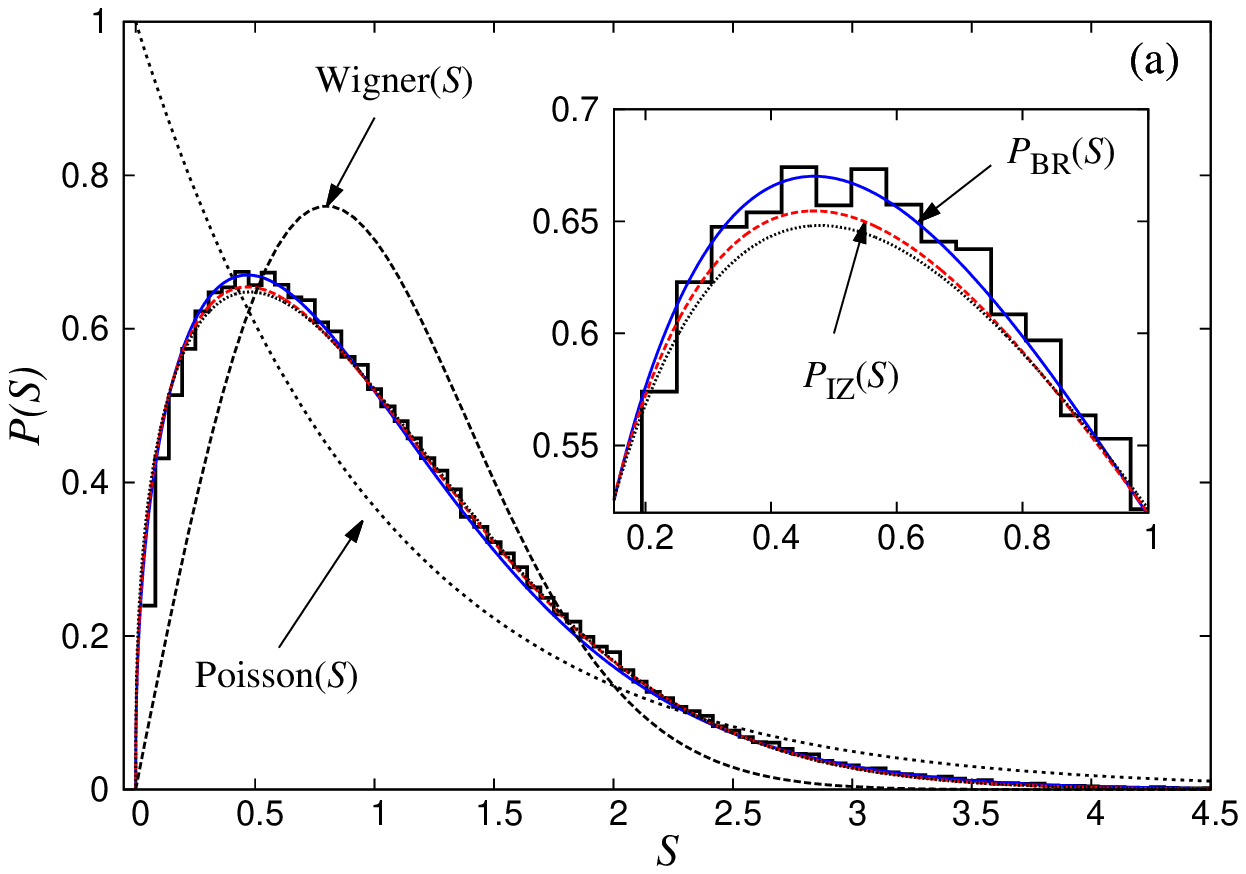}
\includegraphics[width=7.5cm]{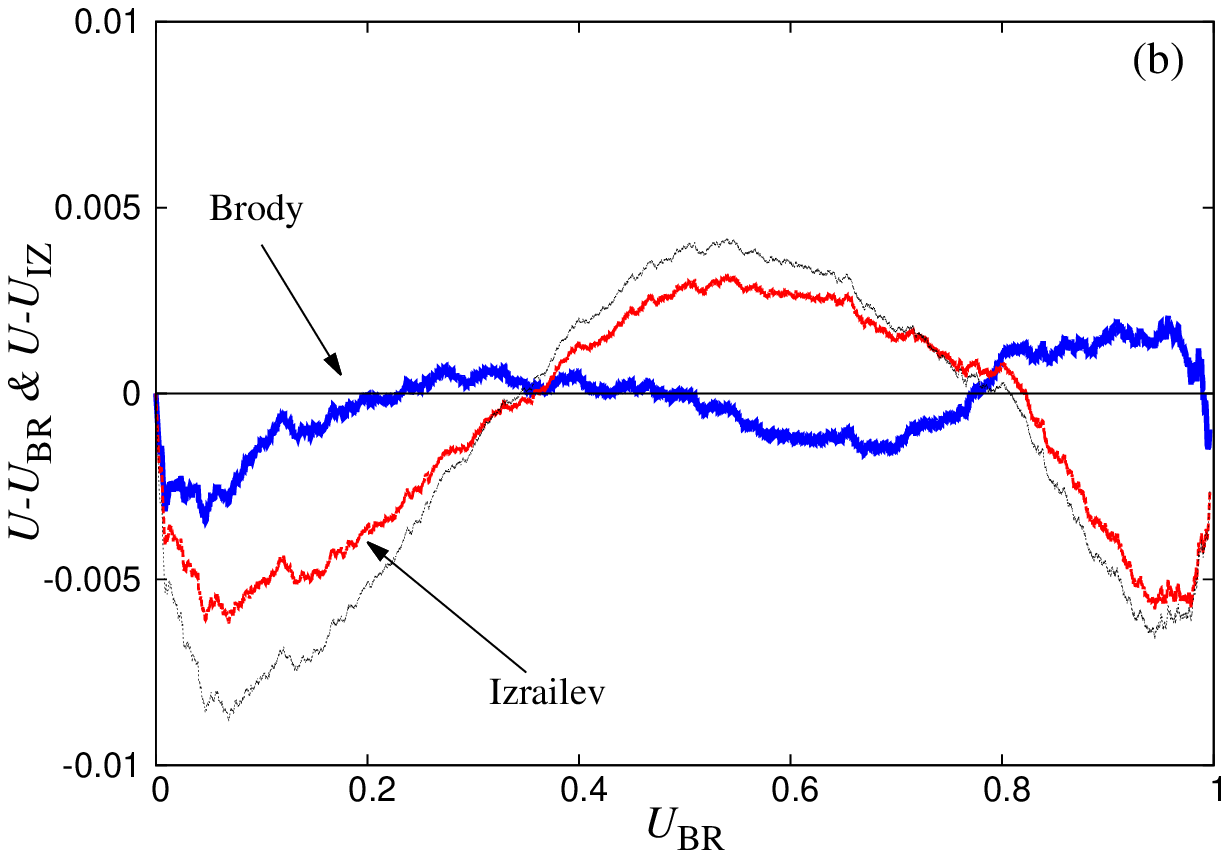}\\
\includegraphics[width=7.5cm]{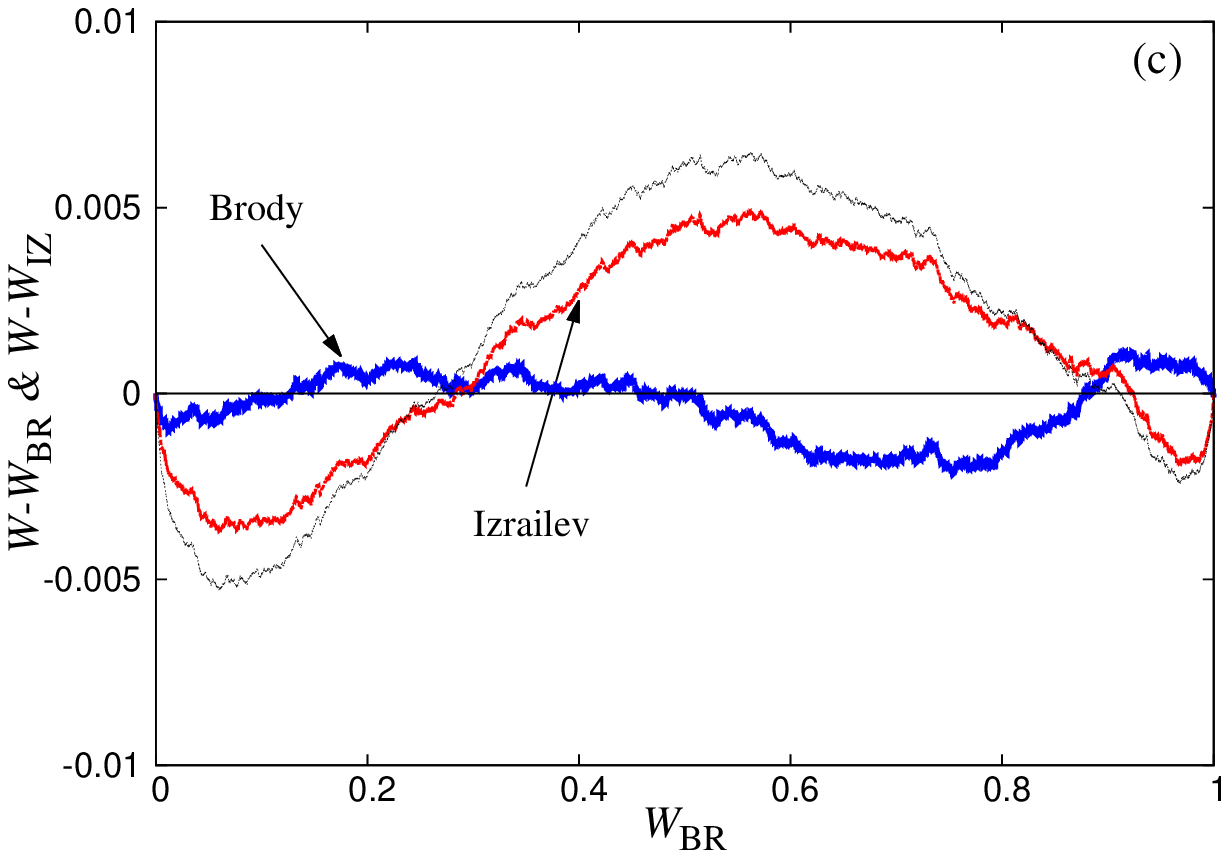}
\includegraphics[width=7.5cm]{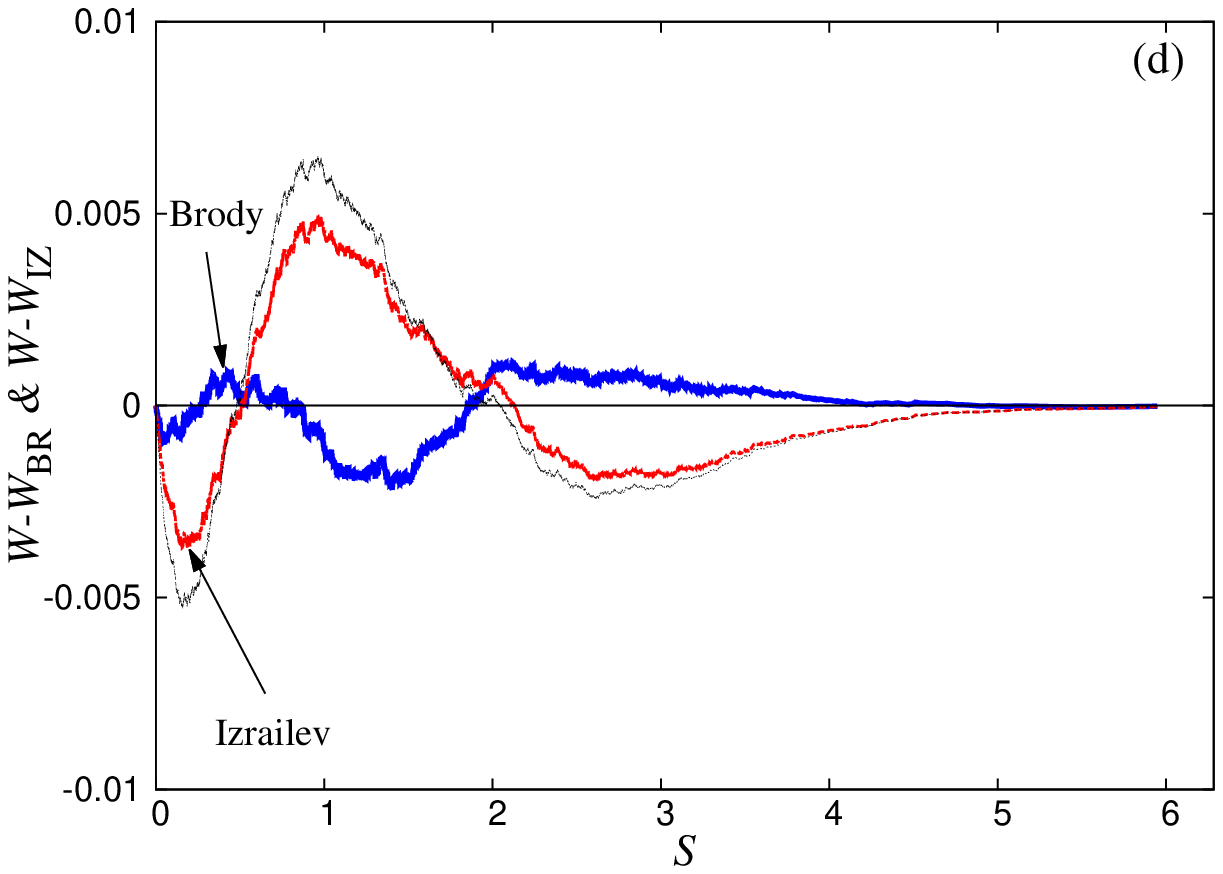}
\caption{Intermediate statistics (panel (a)) for distribution $P(S)$ (histogram - black line) of the model fitted with distribution $P_{\rm BR}$ (blue-solid line), $P_{\rm IZ}$ (red-dashed line pointed by the arrow in the inset figure) and $P_{\rm IZ}^{\rm new}$ (black-dotted line) for $M \times N=641\times398$, $K=7$ and $k=11$ (see text for discussion). In panels (b),(c),(d) we show the difference of the numerical data and the best fitting Brody (blue-thick line), ``old" Izrailev (red-medium line) and Izrailev ``new/improved" (black-thin line; always the outer one) PDFs by using the $U$-function and $W$-distribution. Thus, in case of the ideal fitting the data would lie on the abscissa. In this case, based on the $P(S)$ fit we get $\beta_{\rm BR}$=0.421, $\beta_{\rm IZ}$(old)=0.416 and $\beta_{\rm IZ}$(new)=0.376, and based on the $W(S)$ fit we get $\beta_{\rm BR}$=0.421, $\beta_{\rm IZ}$(old)=0.401, $\beta_{\rm IZ}$(new)=0.350.}
\label{fig5}
\end{figure*}
In Fig.~\ref{fig5}a we show the histogram. It is seen that Brody distribution is better fitting the data than the Izrailev distribution. Since the deviations are really small, the statistical significance very high, we plot in
Fig.~\ref{fig5}b the differences $U({\rm data})-U({\rm Brody/Izrailev})$ versus  $U_{\rm BR}$. Thus if data are on the abscissa the agreement is perfect. As can be seen, the deviations from that are really small, and clearly smaller for Brody. In Fig.~\ref{fig5}c we show the fine differences of $W({\rm data})-W({\rm Brody/Izrailev})$ versus  $W_{\rm BR}$, and again we clearly see that Brody is better. Finally, in Fig.~\ref{fig5}d we show the same thing as in Fig.~\ref{fig5}c, except not versus $W$ but versus $S$ instead. It must be emphasized that the improved Izrailev distribution
(\ref{NewIzr}) exhibits even larger deviations from the data than the original
one (\ref{IzrailevDistrib}), as depicted in Fig.~\ref{fig5} by the outer most line in panels (b,c,d). This is the main reason why we have not considered the new Izrailev distribution (\ref{NewIzr}) any further, and also in order to be compatible with the previous related results in the literature.

More data for various $k$ are shown in Table \ref{tb1}. The index PDF or CDF means that the fitting was done by using $P(S)$ or $W(S)$, respectively. The tick marks indicate which fitting is statistically better, based on the $\chi^2$ procedure. In order to illustrate some more cases from the Table \ref{tb1}, namely for other values of $\beta$, we show the results in Appendix \ref{ap:B}. Even more data are in the Table \ref{tb2}.

Finally, we do a similar analysis for large matrices $N=4000$, and take $M=9$ of them in an ensemble. The results in Appendix \ref{ap:B} clearly show that Brody distribution is an excellent fit to the level spacing distribution in all three representations, $P(S)$, $W(S)$ and $U(S)$.

\subsection{Residues and $\chi^2$ test \label{subsec4.3}}

In the best fitting procedure we have calculated both the residues and the $\chi^2$ as follows:
\begin{itemize}
  \item PDFs residues
  \be\label{Res}
    R_{\rm PDFs}=\sum_{i=1}^{N}(P_{\rm BR,IZ}(i) - {\rm data}(i))^2
  \ee
 \item $\chi^2$
  \be\label{Res}
    \chi^2_{\rm PDFs}=\sum_{i=1}^{N}\frac{(P_{\rm BR,IZ}(i) - {\rm data}(i))^2}{P_{\rm BR,IZ}(i)}.
  \ee
\end{itemize}

\begin{figure*}
\center
\includegraphics[width=5.5cm]{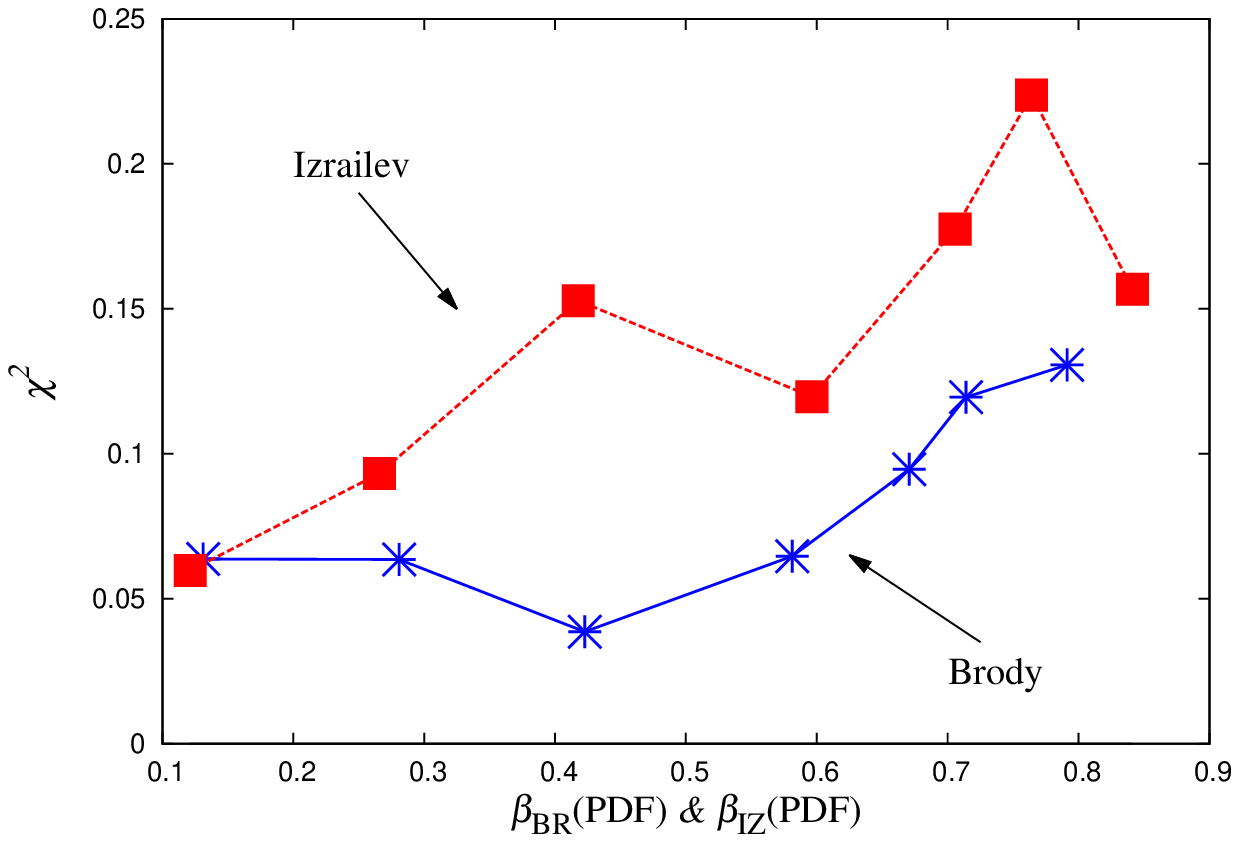}
\includegraphics[width=5.5cm]{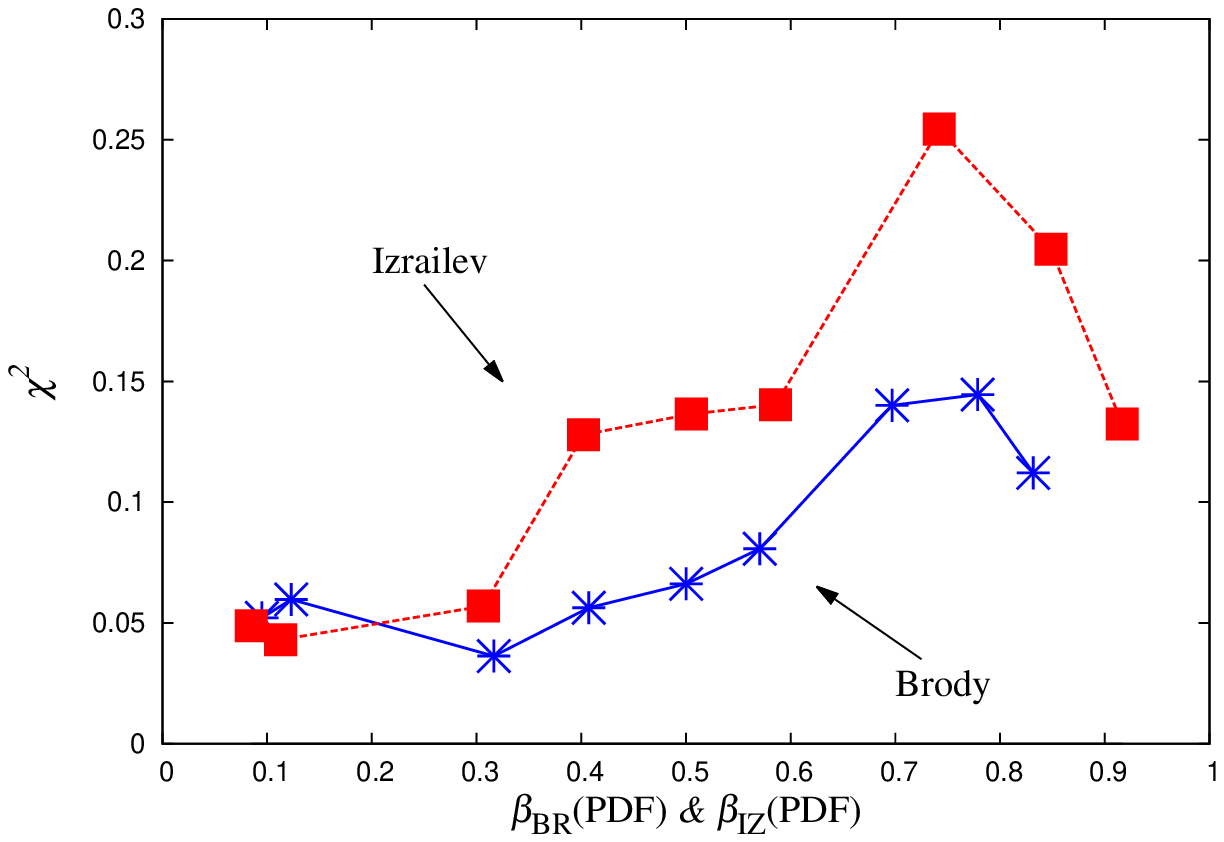}
\includegraphics[width=5.5cm]{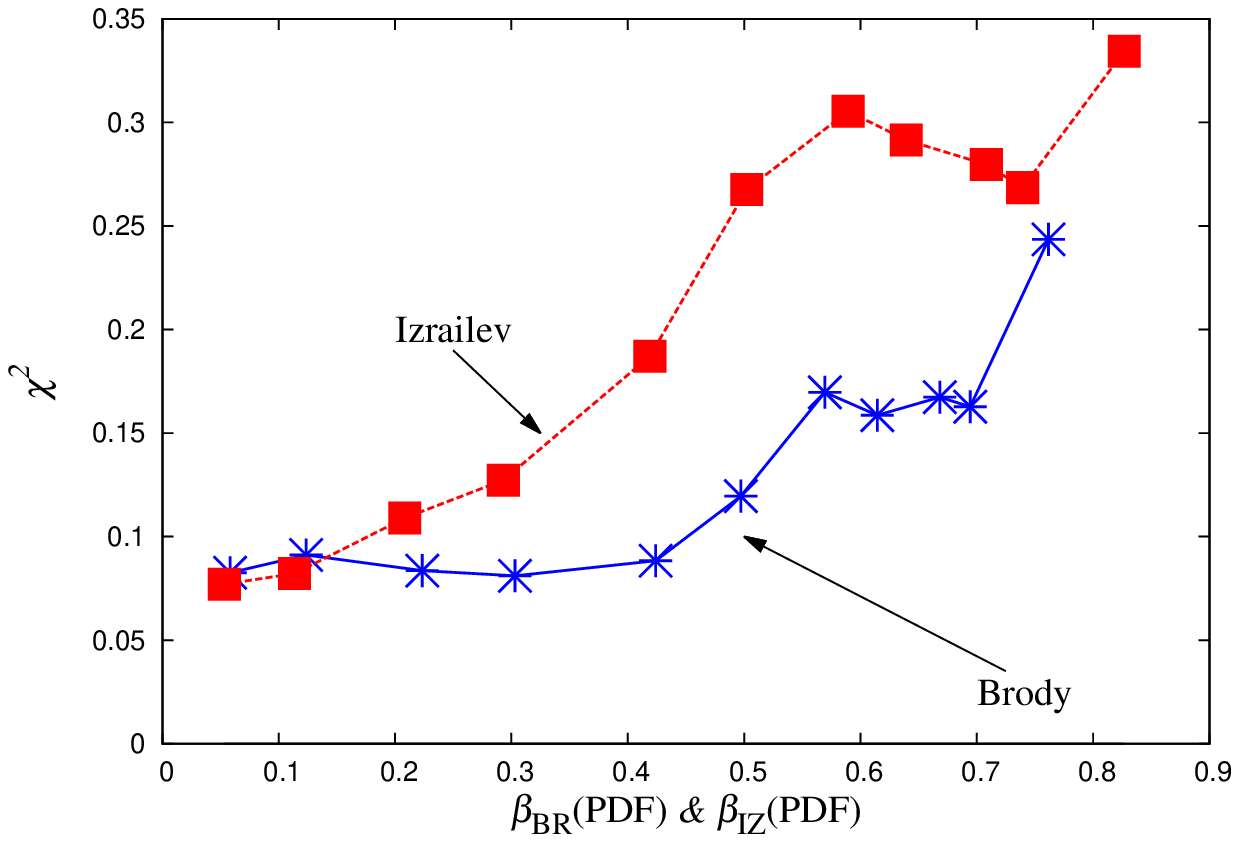}
\caption{$\chi^2$ for $P_{\rm BR}$ and $P_{\rm IZ}$ for $N=398$ (a), $N=796$ (b) and $N=4000$ (c) for various values of the best fits for $\bro$ (black line) and $\biz$ (gray line) using 150 subintervals in $S$.}
\label{fig12}
\end{figure*}

In Fig.~\ref{fig12} we show three examples for the $\chi^2$ as a function of the fitted parameter $\bro$ or $\biz$ for the data from the Tables
\ref{tb1}-\ref{tb3}. It is clearly demonstrated that Brody fit is significantly better than Izrailev.
\begin{table}\caption{\label{tb1}
Results of the best fitting procedure  using Brody (BR) and Izrailev (IZ) distribution. The index PDF or CDF means that the fitting was done by using $P(S)$ or $W(S)$. The tick marks indicate which fitting (BR or IZ) is statistically better, based on the $\chi^2$ procedure.}
\begin{ruledtabular}
\begin{tabular}{cccccc}
\multicolumn{5}{c}{$K=7 \ {\rm and} \ M\times N=161\times 398$} \\
$k$ & $\bro^{\rm PDF}$ & $\bro^{\rm CDF}$ & $\biz^{\rm PDF}$ & $\biz^{\rm CDF}$\\
\colrule
5  & 0.131197 & 0.139808 & 0.121197$\checkmark$ & 0.109808$\checkmark$ \\
8  & 0.280887$\checkmark$ & 0.286922$\checkmark$ & 0.265887 & 0.256920 \\
11 & 0.421398$\checkmark$ & 0.420996$\checkmark$ & 0.416398 & 0.400996 \\
14 & 0.581301$\checkmark$ & 0.574325$\checkmark$ & 0.596301 & 0.574325 \\
17 & 0.670686$\checkmark$ & 0.659891$\checkmark$ & 0.705686 & 0.679891 \\
20 & 0.713984$\checkmark$ & 0.699906$\checkmark$ & 0.763984 & 0.739906 \\
23 & 0.791288$\checkmark$ & 0.778780$\checkmark$ & 0.861288 & 0.838780 \\
26 & 0.812993$\checkmark$ & 0.797852$\checkmark$ & 0.892993 & 0.867852 \\
29 & 0.832067$\checkmark$ & 0.821543 & 0.912067 & 0.891543$\checkmark$ \\
\end{tabular}
\end{ruledtabular}
\end{table}
\begin{table}\caption{\label{tb2}
Results of the best fitting procedure. The same as in Table \ref{tb1}, but for  $N=796$ and other parameter values $k$.}
\begin{ruledtabular}
\begin{tabular}{cccccc}
\multicolumn{5}{c}{$K=7 \ {\rm and} \ M\times N=81\times 796$} \\
$k$ & $\bro^{\rm PDF}$ & $\bro^{\rm CDF}$ & $\biz^{\rm PDF}$ & $\biz^{\rm CDF}$\\
\colrule
6    & 0.094946 & 0.102980 & 0.084946$\checkmark$ & 0.082980$\checkmark$ \\
7    & 0.122990 & 0.135550 & 0.11299$\checkmark$ & 0.1079550$\checkmark$ \\
12.5 & 0.316602$\checkmark$ & 0.319169$\checkmark$ & 0.306602 & 0.289169 \\
15   & 0.407302$\checkmark$ & 0.400864$\checkmark$ & 0.402302 & 0.380864 \\
17.5 & 0.500219$\checkmark$ & 0.491786$\checkmark$ & 0.505219 & 0.481786 \\
20   & 0.570621$\checkmark$ & 0.560884$\checkmark$ & 0.585621 & 0.560884 \\
25   & 0.696955$\checkmark$ & 0.680933$\checkmark$ & 0.741955 & 0.710933 \\
30   & 0.778601$\checkmark$ & 0.761853$\checkmark$ & 0.848601 & 0.821853 \\
35   & 0.831750$\checkmark$ & 0.816494$\checkmark$ & 0.911750 & 0.886494 \\
\end{tabular}
\end{ruledtabular}
\end{table}
\begin{table}\caption{\label{tb3}
Results of the best fitting procedure. The same as in Table~\ref{tb1}, but for $N=4000$ and other parameter values $k$.}
\begin{ruledtabular}
\begin{tabular}{cccccc}
\multicolumn{5}{c}{$K=7 \ {\rm and} \ M\times N=9\times 4000$} \\
$k$ & $\bro^{\rm PDF}$ & $\bro^{\rm CDF}$ & $\biz^{\rm PDF}$ & $\biz^{\rm CDF}$\\
\colrule
10 & 0.058254 & 0.064513 & 0.053254$\checkmark$ & 0.055413$\checkmark$ \\
15 & 0.123504 & 0.136268 & 0.113504$\checkmark$ & 0.116268$\checkmark$ \\
20 & 0.223351$\checkmark$ & 0.228405$\checkmark$ & 0.208351 & 0.198405 \\
25 & 0.303080$\checkmark$ & 0.304349$\checkmark$ & 0.293080 & 0.274349 \\
30 & 0.424033$\checkmark$ & 0.410826$\checkmark$ & 0.419033 & 0.390826 \\
35 & 0.497268$\checkmark$ & 0.485010$\checkmark$ & 0.502268 & 0.475010 \\
40 & 0.569422$\checkmark$ & 0.553461$\checkmark$ & 0.586422 & 0.553461 \\
45 & 0.614519$\checkmark$ & 0.596811$\checkmark$ & 0.639519 & 0.606811 \\
50 & 0.668207$\checkmark$ & 0.651155$\checkmark$ & 0.708207 & 0.671155 \\
55 & 0.694437$\checkmark$ & 0.679703$\checkmark$ & 0.739437 & 0.709703 \\
60 & 0.761739$\checkmark$ & 0.740879$\checkmark$ & 0.821739 & 0.790879 \\
\end{tabular}
\end{ruledtabular}
\end{table}

\section{The scaling behaviour of $\blo$ versus $\Lambda$  \label{sec5}}

\begin{figure}
\center
\includegraphics[width=7.5cm]{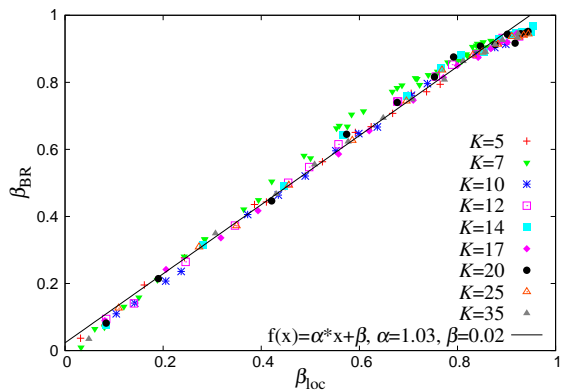}
\caption{[Colour online] The fit parameter $\bro$ as a function of $\blo$ for $161\times398$ elements for various values of $K$ and for a wide range of $k$ values. The best fitting straight line is very close to identity.}
\label{fig10}
\end{figure}

\begin{figure}
\center
\includegraphics[width=7.5cm]{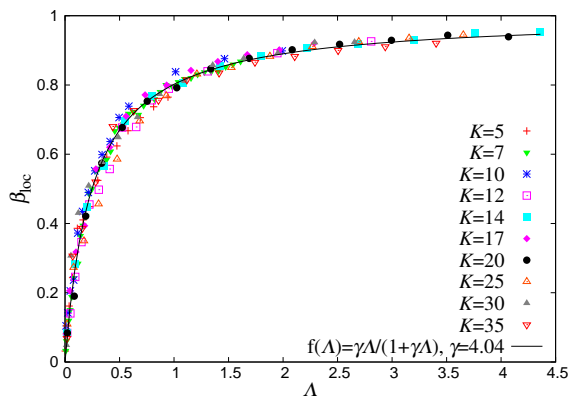}
\caption{ [Colour online] The parameter $\blo$ vs. $\La$ for various $K$ and $k$, (the same as in Fig.~\ref{fig10}) where the scaling law [Eq.~(\ref{bla})] is shown with the black line.}
\label{fig13}
\end{figure}

It is well established that the degree of localization and the value of the spectral level repulsion parameter $\beta$  are related. The parameter that determines the transition from weak to strong quantum chaos is neither the strength parameter $k$ nor the localization length $l_{\infty}$, but the ratio of the localization length $l_{\infty}$ to the size $N$ of the system in momentum $p$, the scaling parameter $\La$ [Eq.~(\ref{MLL:Lamda})]. Here we present the following results
\begin{itemize}
\item In plotting $\bro$ versus $\blo$ as shown in Fig.~\ref{fig10} for a great number of matrices at various parameter values we clearly see the linear relationship very close to identity.
\item In Fig.~\ref{fig13} we plot $\blo$ versus $\La$ for the same ensemble of matrices as in Fig.~\ref{fig10}:  clearly, there is a functional relationship according to the scaling law
    \be \label{bla}
    \blo = \frac{\gamma \La}{1 +\gamma \La},\;\;\; \gamma=4.04,
    \ee
    which is similar to the scaling law [Eq.~(\ref{beta_loc_scaling})], but not the same. Thus we see that when $\La\rightarrow \infty$ both $\blo$ and $\bro$ go to one, and we have extended eigenstates and GOE/COE spectral statistics, whilst in the limit $\La =0$ we have strong localization, $\bro$ and $\blo$ are zero, and we have Poissonian spectral statistics. Fig.~\ref{fig10} shows what happens in between. The value $\gamma=4.04$ differs somewhat from $\gamma\approx 3.2$ in \cite{CGIS1990}, where $\blo$ is plotted versus $x\approx 4\La$.

\item The quality of the fit of Fig.~\ref{fig10} is degraded a lot when the size of the ensemble is decreased, as we have observed.

\item The quality of the fit when using $\biz$ instead of $\bro$ is decreased, as we have checked carefully and as shown in Tables \ref{tb1}, \ref{tb2} and \ref{tb3}.
\end{itemize}

The numerical factor $\alpha_{\mu}$ is determined by numerical calculations, namely by seeking the best agreement in Fig.~\ref{fig13}. It is interesting to note that in the case $K=5$ extensively studied by Izrailev (\cite{Izr1990} and references therein) we find $\mu=0.99$, which is compatible with $\mu=1$, but $\alpha_{\mu}=0.45$ which is only approximately his value  $0.5$. Then $D_1\approx K^2/2$ and $K^2/\tau^2 =k^2$ and we have $\linf \approx k^2/4$. The data for all cases that we treat in this paper are given in Table~\ref{tb4}.

In \cite{CGIS1990} the following scaling law was proposed
\be \label{beta_loc_scaling}
    \blo=\frac{\gamma x}{1+\gamma x}, \ {\rm where} \ x\equiv \frac{k^2}{N} \ {\rm and} \ \gamma \approx 0.8.
\ee
A banded random matrix model has been proposed \cite{CMI1990,CIM1991,CGIFM1992,CCGI1993} to explain the above scaling relationship, based on the fact that Eq.~(\ref{Fmatrix}) is an infinite banded matrix of band width approximately $k$, which can be reduced to the finite dimensional model with the same property. However, one should observe the fact that $k^2/(4N)$ is just an approximate value of $\Lambda$ valid for the special case when $\mu=1$ (normal diffusion), $\alpha_1=1/2$ and $K=5$, where $D_1\approx K^2/(2\tau^2)= k^2/2$. In fact, we find numerically $\alpha_1=0.45$, not $0.5$. Moreover, we must be aware of the fact that at $K=5$ we still have islands of stability in the classical phase space, implying problems with the divided phase space in the quantum picture. This is the reason why we consider the cases $K\ge 7$, but nevertheless check Izrailev's results \cite{Izr1990} limited to $K=5$ (Table~\ref{tb4}). As seen in Fig.~\ref{fig15}, as an example,  in case $K=7$ we have three different diffusion regimes, and for our purposes the middle regime with $\mu\approx 0.9$ and $D_{\mu}\approx 169.82$ is relevant and  important, because $\linf$ is in the range of $N$.
\begin{table}\caption{\label{tb4}
We show the values of $\mu$, $D_{\mu}$ for the classical diffusion. The coefficients $\alpha_{\mu}$ are needed for the quantum kicked rotator in estimating $\linf$, and they are also determined numerically by seeking the best agreement in Fig.~\ref{fig13}.}
\begin{ruledtabular}
\begin{tabular}{cccc}
$K$ & $\mu$ & $\alpha_{\mu}$ & $D_{\mu}$ \\
\colrule
5 & $\approx$ 0.99 & 0.45 & 13.182  \\
7 & $\approx$ 0.90 & 0.20 & 169.82  \\
10 & $\approx$ 1.00 & 0.50 & 31.68  \\
12 & $\approx$ 1.00 & 0.50 & 87.09  \\
14 & $\approx$ 1.00 & 0.50 & 134.89 \\
17 & $\approx$ 1.01 & 0.50 & 97.05  \\
20 & $\approx$ 0.99 & 0.50 & 275.42 \\
25 & $\approx$ 0.98 & 0.50 & 405.50 \\
30 & $\approx$ 1.00 & 0.50 & 389.04 \\
35 & $\approx$ 1.00 & 0.50 & 467.73 \\
\end{tabular}
\end{ruledtabular}
\end{table}
In the present work we thus generalize the empirical scaling law [Eq.~(\ref{bla})] to all values of $K\ge K_{crit} \approx 0.97$ and show that the scaling $\blo$ versus $\Lambda$ persists under the proper generalizations, now including anomalous diffusion.

Thus the knowledge of the theoretical scaling parameter $\La$ from Eq.~(\ref{MLL:Lamda}), in the semiclassical limit, enables us to calculate the spectral level repulsion parameter $\bro$ (or $\biz$) according to the scaling law [Eq.~(\ref{bla})]. This law clearly shows that at fixed parameters $K$, $\tau$ and $k=K/\tau$ in the limit $N\rightarrow \infty$ we always get $\bro=0$, i.e. Poisson distribution. On the other hand, when $\linf$ becomes greater than $N$ we see approach to the random matrix (GOE or COE) behaviour.
The scaling properties of this section are one of the main conclusions of this paper. Recently, strong evidence has been found that a similar relationship can exist for the analogous quantities in dynamically localized time-independent Hamiltonian systems \cite{BatRob2010,BatRob2012}.

\section{Discussion and conclusions  \label{sec6} }

In this paper we study in detail the relationship between the localization measure of the eigenstates and the spectral level repulsion parameter in the quantum kicked rotator. First we confirm and improve the results of references \cite{Izr1986,Izr1987,Izr1988,Izr1989,Izr1990} for $K=5$, and then we go substantially beyond his work by doing the analysis for many different values of $K\ge 5$ (dimensionless classical kick parameter), not only $K=5$, namely $K\in[5,35]$, with various classical dynamical regimes,  and many different $k\in [5,60]$ (dimensionless quantum kick parameter). Namely, we include also the cases with accelerator modes and generalized (anomalous) diffusion (subdiffusion and superdiffusion).

The classical kicked rotator [Eq.~(\ref{KR})] is one of the most important model systems in classical and quantum chaos. We have studied the semiclassical regime where $k\ge K$ and analyzed the eigenstates and eigenvalues (quasienergies, or eigenphases $\phi_n\in[0,2\pi)$), in particular the aspects of the dynamical localization. The infinite dimensional case has a finite localization length $\linf$ (in the space of the angular momentum quantum numbers), and exhibits the Poissonian statistics of the level spacings. We study a finite $N$-dimensional model [Eqs.~(\ref{u_repres}-\ref{Bnmoper})] proposed by Izrailev \cite{Izr1986,Izr1987,Izr1988,Izr1989,Izr1990} for the case of odd-parity eigenstates. In this model the intermediate spectral statistics is observed, ranging from Poissonian to the RMT statistics (GOE/COE statistics), depending on the value of the scaling parameter $\La = \linf/N$. We have shown that $\La$ can be calculated theoretically in terms of generalized classical diffusion properties of the standard map expressed as $l_{\infty}= \left( \frac{\alpha_{\mu} D_{\mu} (K)}{\tau^2}\right)^{\frac{1}{2-\mu}}$ in Eq.~(\ref{finallinf}), which is a major result of this paper as the anomalous diffusion ($\mu\not=1$) is now included. The degree of localization is also estimated by means of information entropy described by the parameter $\blo$, which goes from $0$ (Poisson) to $1$ (GOE), and is uniquely determined by $\La$, as described by the scaling law $\blo = \frac{\gamma \La}{1 +\gamma \La},\;\;\; \gamma=4.04$ in Eq.~(\ref{bla}).

We find that in almost all cases the Brody distribution correctly captures the level spacing distribution at all values of the corresponding level repulsion exponent $\bro$, noticeably better than the distribution function proposed in \cite{Izr1988,Izr1989}, and also in \cite{CIM1991}, as demonstrated in section \ref{sec4}. We confirm and significantly refine the result that $\bro$ is identical to $\blo$ within the statistical fluctuations.

These results have been obtained for the time-dependent system, the quantum kicked rotator, but we have evidence that similar conclusions can be reached in time-independent chaotic Hamiltonian systems, either in the mixed type regime
or fully chaotic regime, when studying (after separation of regular and chaotic eigenstates) the localized chaotic eigenstates and the statistics of the corresponding chaotic (irregular) level sequences. This has been confirmed in the case of the billiard with mixed type dynamics \cite{Rob1983,Rob1984} in the recent paper by Batisti\'c and Robnik \cite{BatRob2010} indirectly, and directly very recently \cite{BatRob2012,BatManRob2013}. Moreover, it has been shown that different but equivalent localization measures can be introduced which are simply related to the Brody parameter (Batisti\'c and Robnik 2013).
In case of billiards the quality of the spectral statistics is even much
better than in the case of the quantum kicked rotator, due to the possibility to calculate a much larger number of high-lying eigenstates. Another important paradigmatic model is the hydrogen atom in strong magnetic field \cite{Rob1981,Rob1982,HRW1989,WF1989}, which in addition to the various billiards is a good candidate for further theoretical and experimental studies.

To derive the fractional power law level repulsion and the emerging Brody distribution, as a consequence of the dynamical localization in chaotic eigenstates, is an important open theoretical question.

\section*{Acknowledgements}

This work was supported by the Slovenian Research Agency (ARRS). T.M. was also partially supported by a grant from the GSRT, Greek Ministry of Development, for the project ``Complex Matter'', awarded under the auspices of the ERA Complexity Network and by the European Union (European Social Fund) and Greek national funds through the Operational Program ``Education and Lifelong Learning'' of the National Strategic Reference Framework (NSRF) - Research Funding Program: THALES. Investing in knowledge society through the European
Social Fund. Finally, the authors would like to thank B.~Batisti\'c for his fruitful comments and discussions on this work. We would also like to acknowledge the computational facilities of the Center for Research and Applications of Nonlinear Systems (CRANS) of the Department of Mathematics, University of Patras, Greece, where part of the simulations was done.

\appendix

\section{The U-function representation of the level spacing distribution \label{ap:A}}

First we estimate the expected fluctuation (error) of the cumulative (integrated) level spacing distribution $W(S)$, which contains $N_s$ objects.
At a certain $S$ we have the probability $W$ that a level is in the interval $[0,W]$ and $1-W$ that it is in the interval $[W,1]$. Assuming binomial probability distribution $P(k)$ of having $k$ levels in the first and $N_s-k$ levels in the second interval we have
\be \label{A1}
P(k) = \frac{N_s!}{k! (N_s-k)!} W^k (1-W)^{N_s-k}.
\ee
Then the average values are equal to $<k> = N_sW$, $<k^2> = N_sW + N_s(N_s-1) W^2$
and the variance $V(k) = <k^2> - <k>^2  = N_s W (1-W)$. But the probability $W$ is estimated in the mean as $k/N_s$. Its variance is
\be \label{A4}
V(W) = V\left( \frac{k}{N_s} \right) = \frac{1}{N_s^2} V(k) =  \frac{W(1-W)}{N_s}
\ee
and therefore the estimated error of $W$ (standard deviation, the square root of the variance) is given by
\be \label{A5}
\delta W = \sqrt {V(W)} = \sqrt {\frac{W(1-W)}{N_s} }.
\ee
Transforming now from $W(S)$ to
\be \label{A6}
U(S) = \frac{2}{\pi} \arccos \sqrt{1 - W(S)},
\ee
we show in a straightforward manner that $\delta U = \frac{1}{\pi \sqrt{ N_s}}$ and is indeed independent of $S$. From the (choice of the constant pre-factor in the) definition Eq.~(\ref{A6}) one sees that both $U(S)$ and $W(S)$ go from $0$ to $1$ as $S$ goes from $0$ to infinity.

\section{More spectral statistical analysis \label{ap:B}}

\begin{figure*}
\center
\includegraphics[width=6cm]{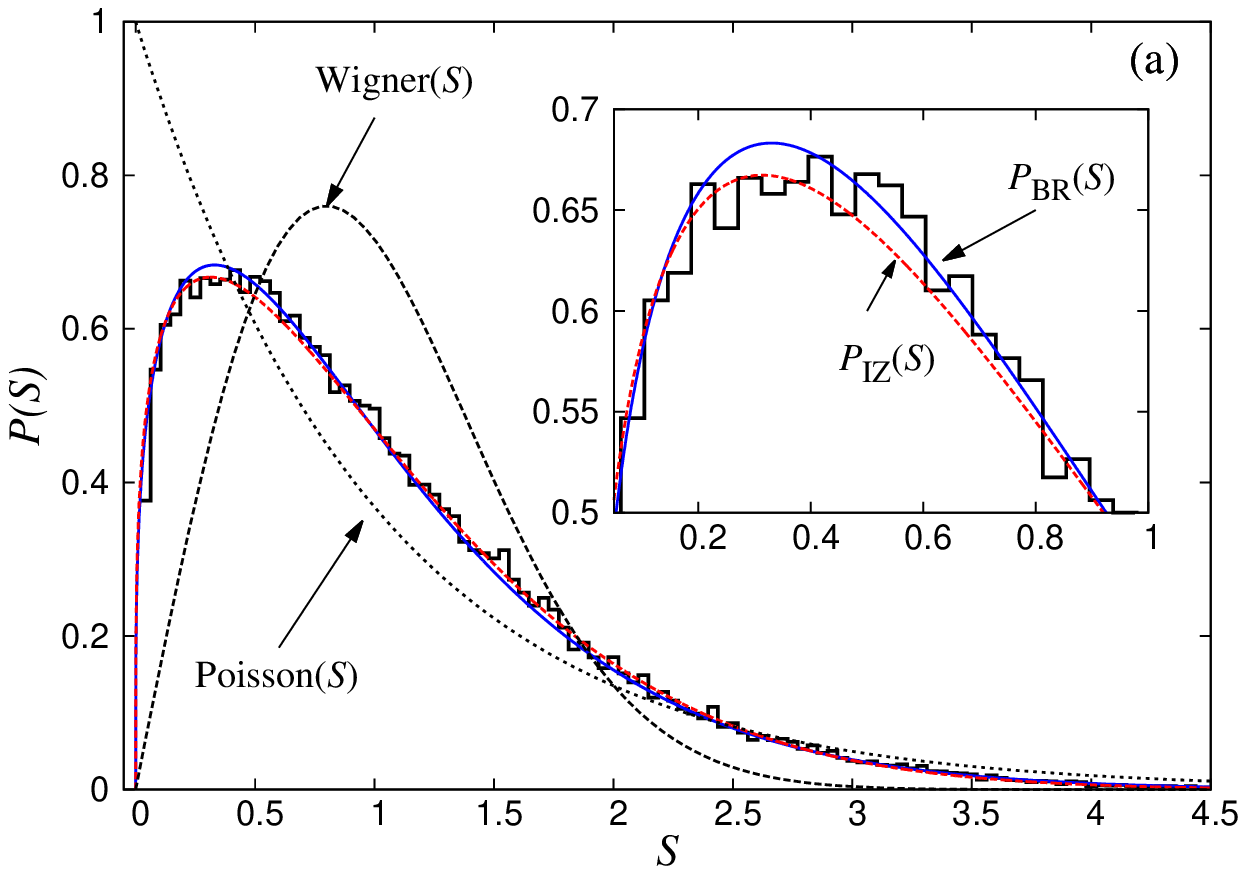}
\includegraphics[width=6cm]{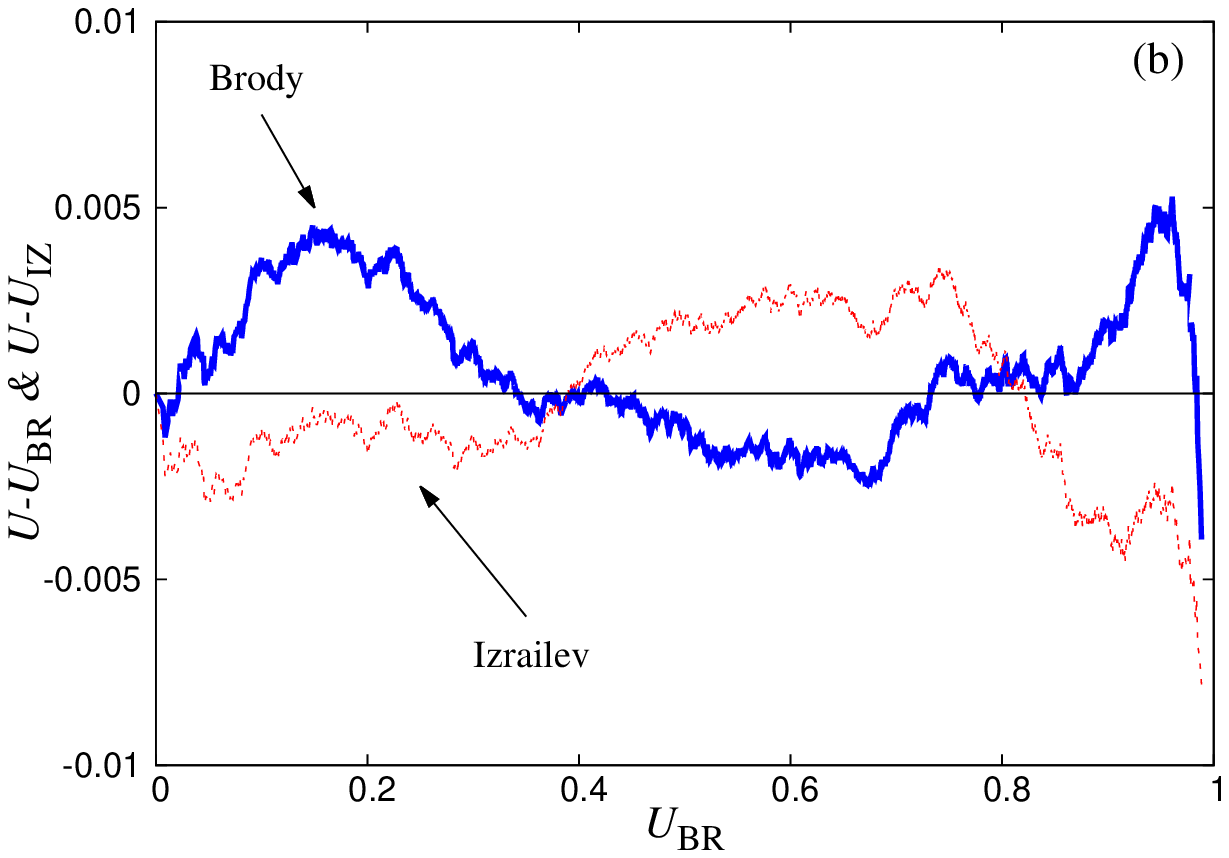}\\
\includegraphics[width=6cm]{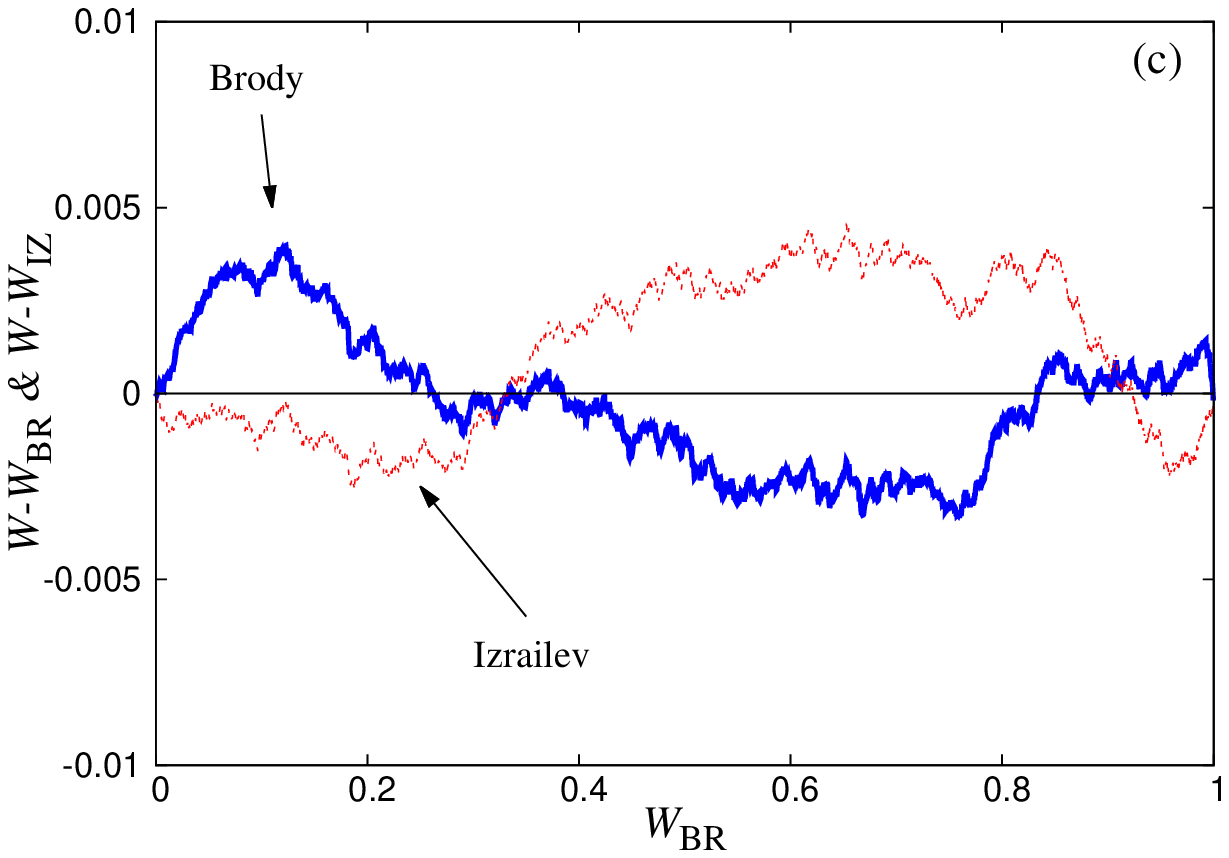}
\includegraphics[width=6cm]{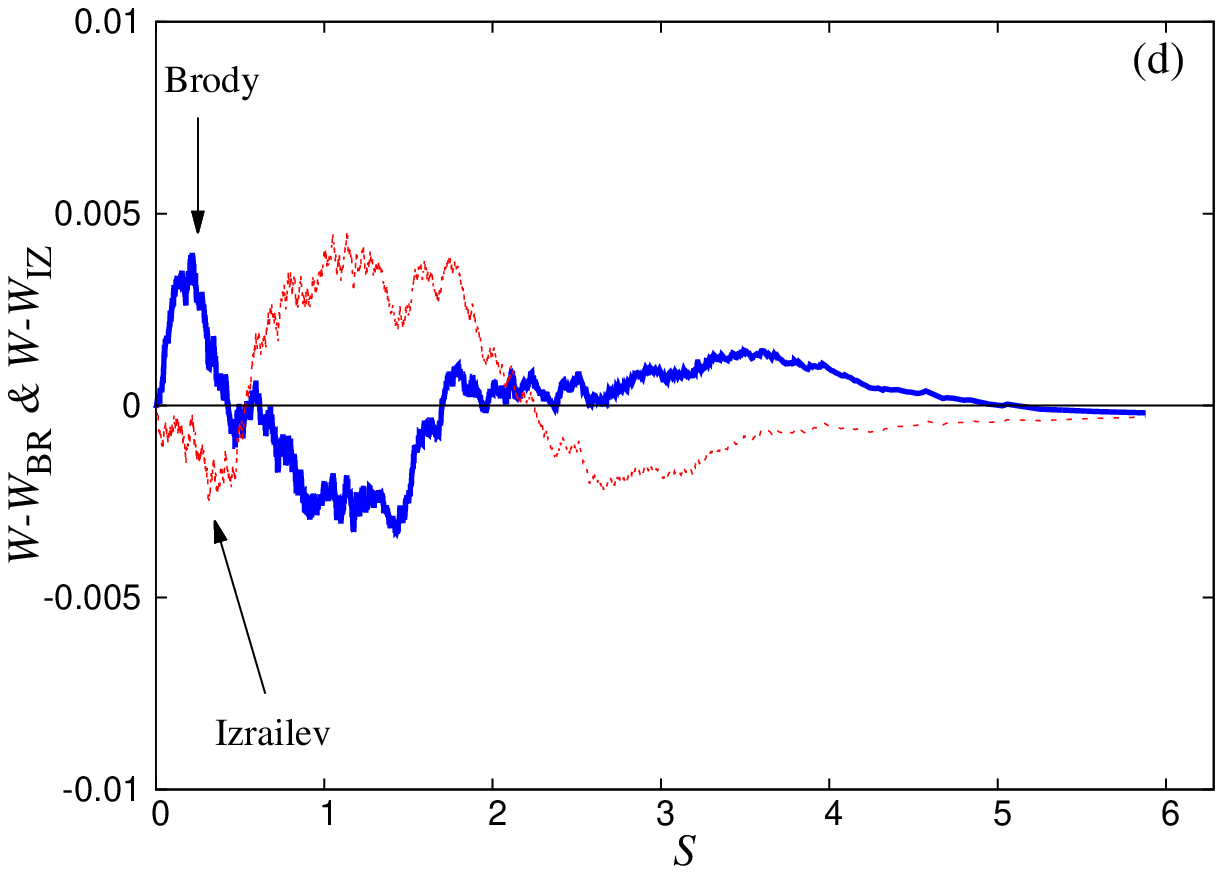}
\caption{Intermediate statistics (panel (a)) for distribution $P(S)$ (histogram black-solid line) of the model [Eqs.~(\ref{u_repres})-(\ref{Unm})] fitted with distribution $P_{\rm BR}(S)$ (blue-solid line) ($\bro=0.28$) and $P_{\rm IZ}(S)$ (red-dashed line) for $M\times N=161\times 398$,  $K=7$ and $k=8$ (see text for discussion). The two black-dotted lines indicate the two extreme distributions, i.e. the Poisson and Wigner. In panels (b),(c),(d) we show the difference of the numerical data and the best fitting Brody (blue-thick-solid line) and Izrailev (red-thin-dashed line) PDFs by using the $U$-function and $W$-distribution (see text for discussion). Thus in case of the ideal fitting the data would lie on the abscissa. See also Table~\ref{tb1}.}
\label{fig7}
\end{figure*}

\begin{figure*}
\center
\includegraphics[width=6.cm]{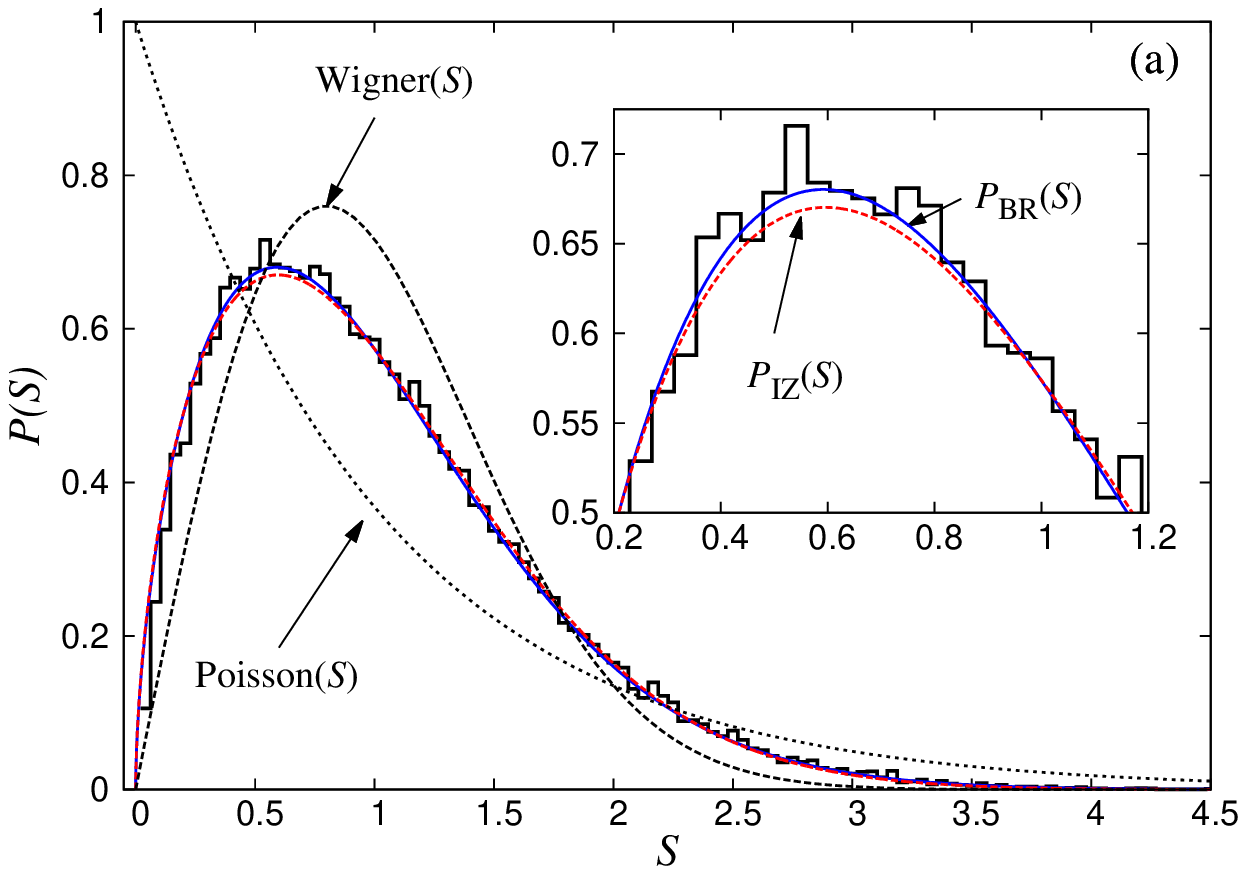}
\includegraphics[width=6cm]{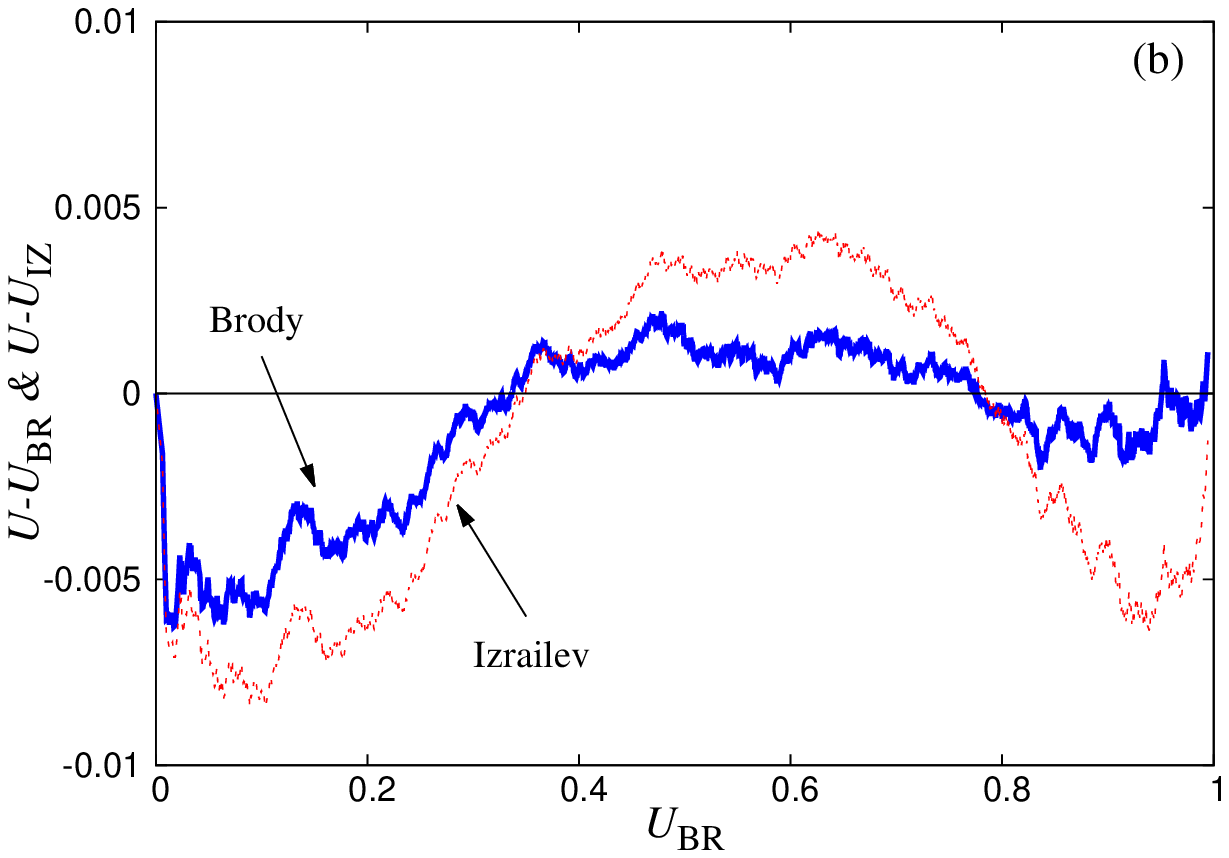}\\
\includegraphics[width=6cm]{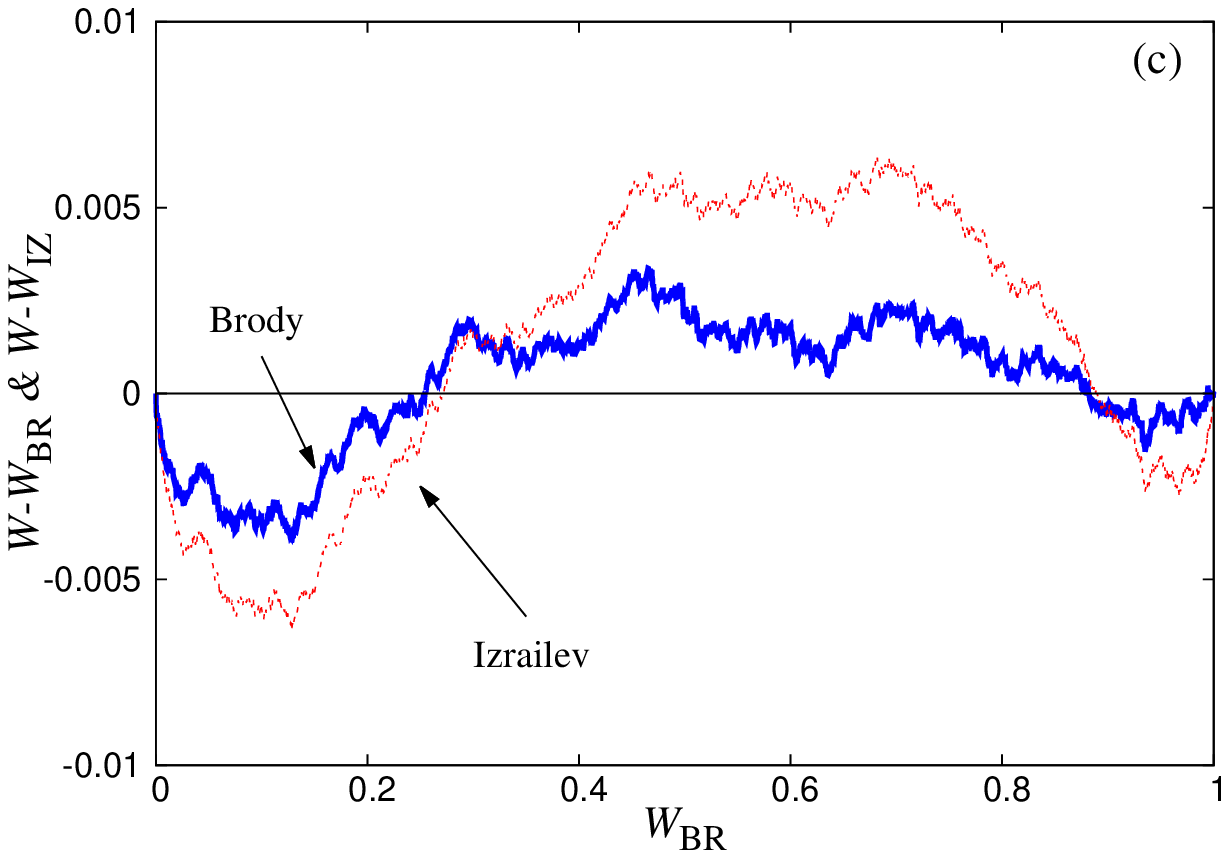}
\includegraphics[width=6cm]{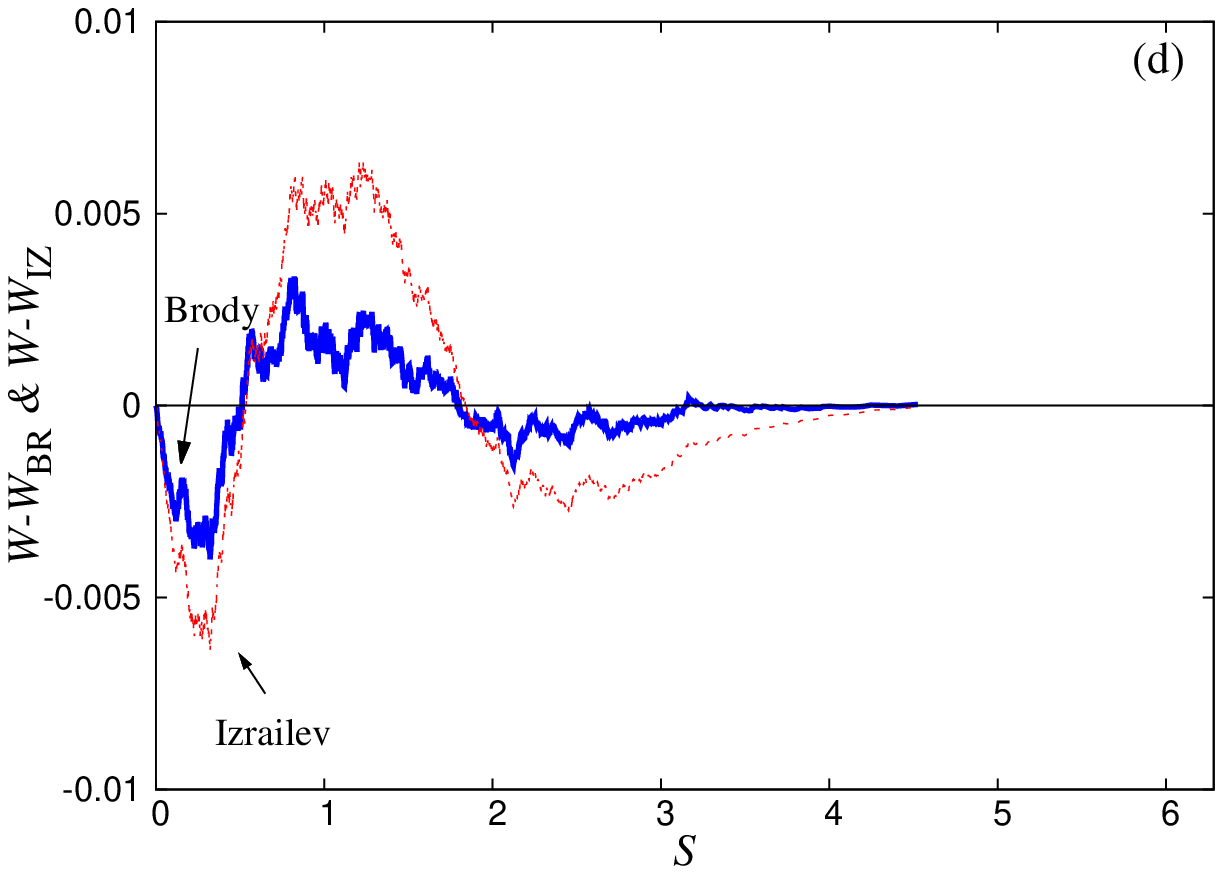}
\caption{Same as Fig.~\ref{fig7} for $M\times N=161\times 398$, $K=7$ and $k=14$ with $\bro=0.58$. See also Table~\ref{tb1} and text for discussion.}
\label{fig8}
\end{figure*}

\begin{figure*}
\center
\includegraphics[width=6cm]{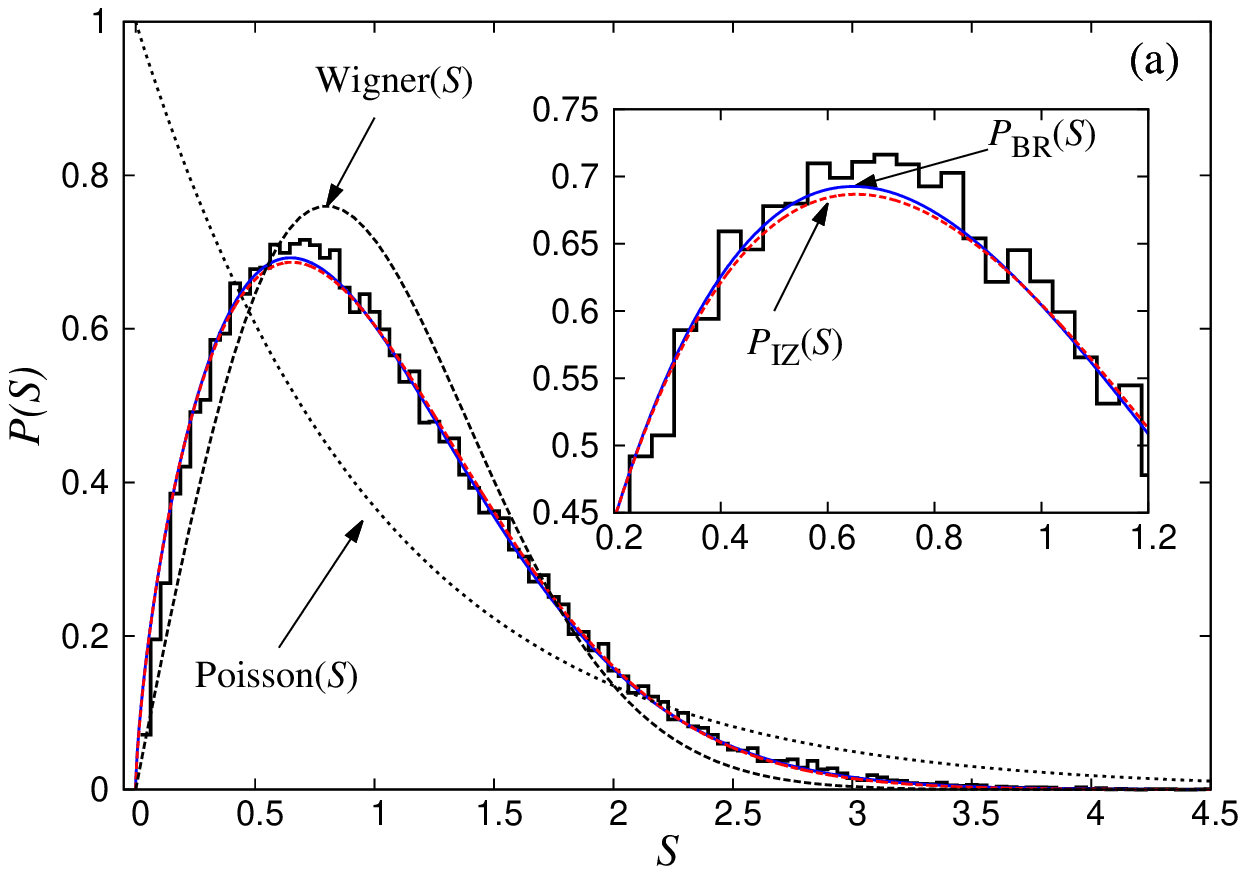}
\includegraphics[width=6cm]{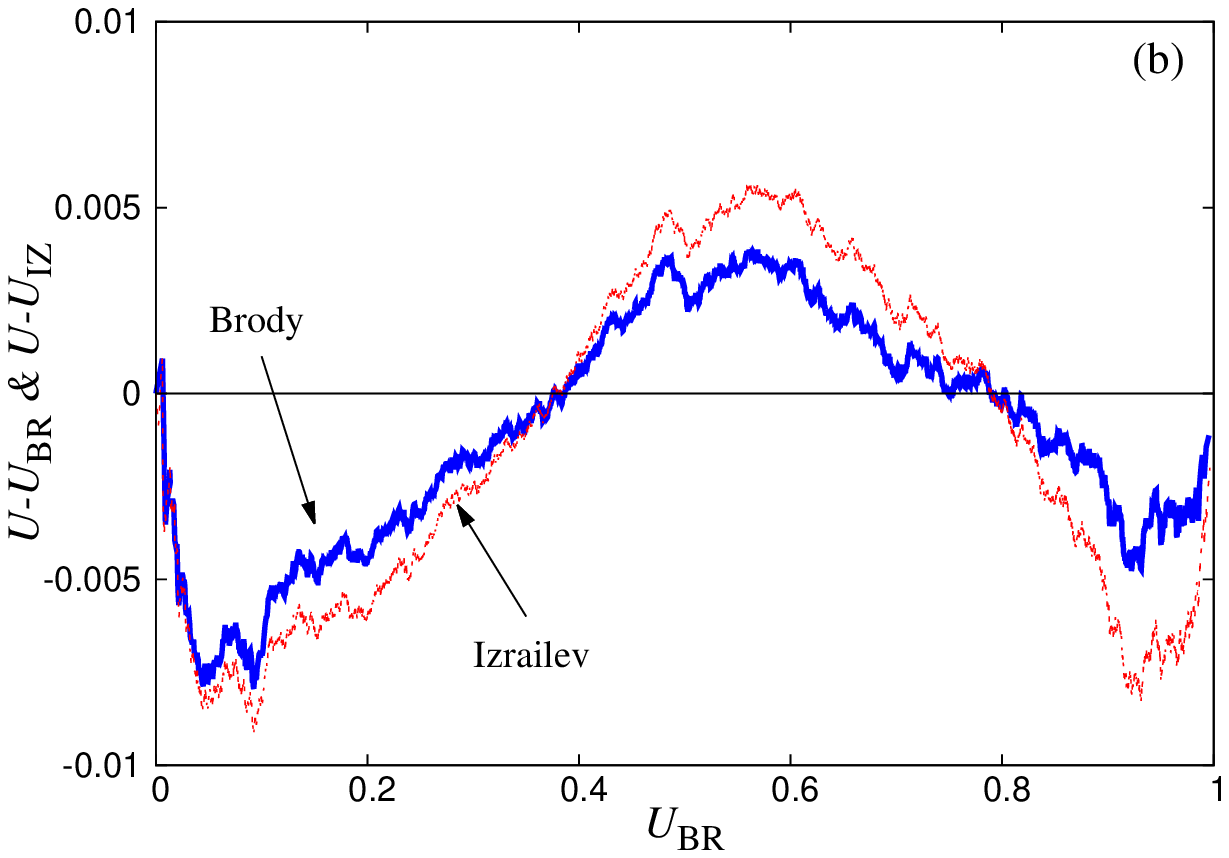}\\
\includegraphics[width=6cm]{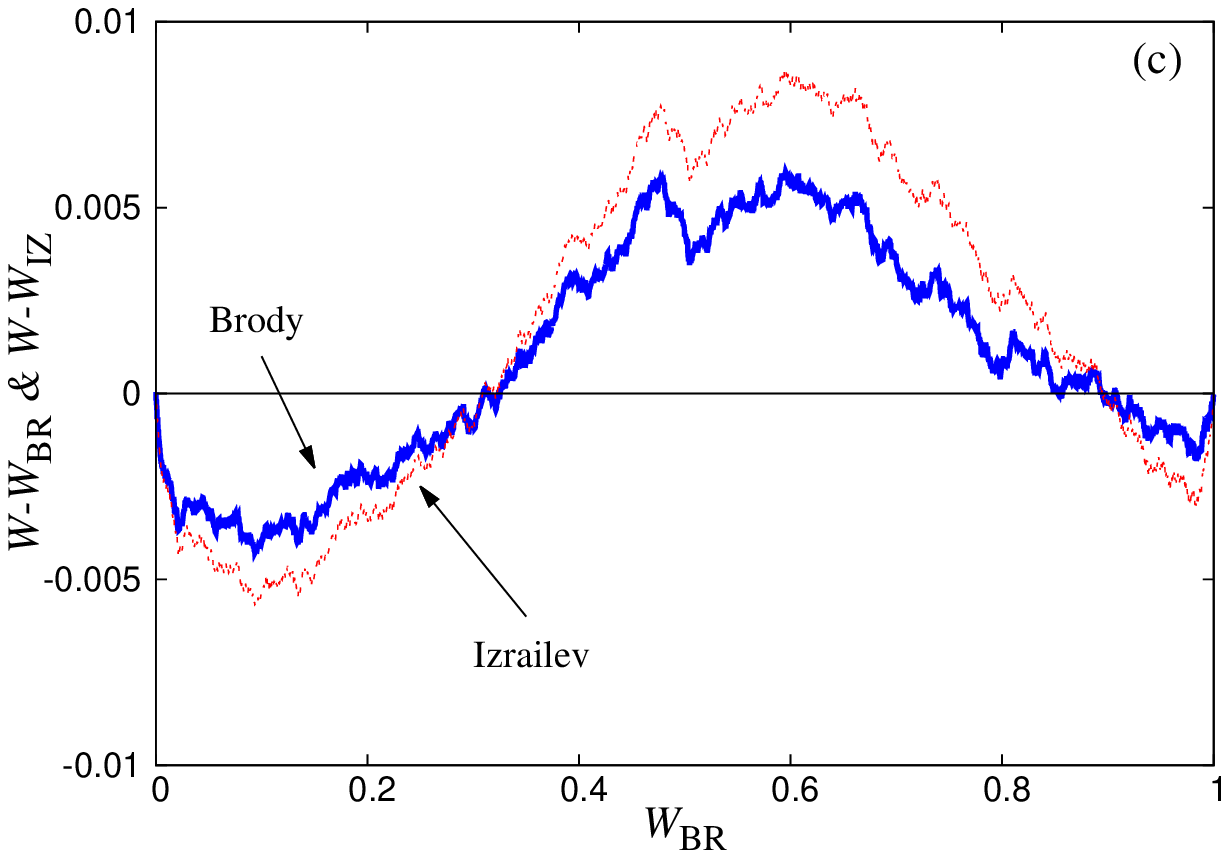}
\includegraphics[width=6cm]{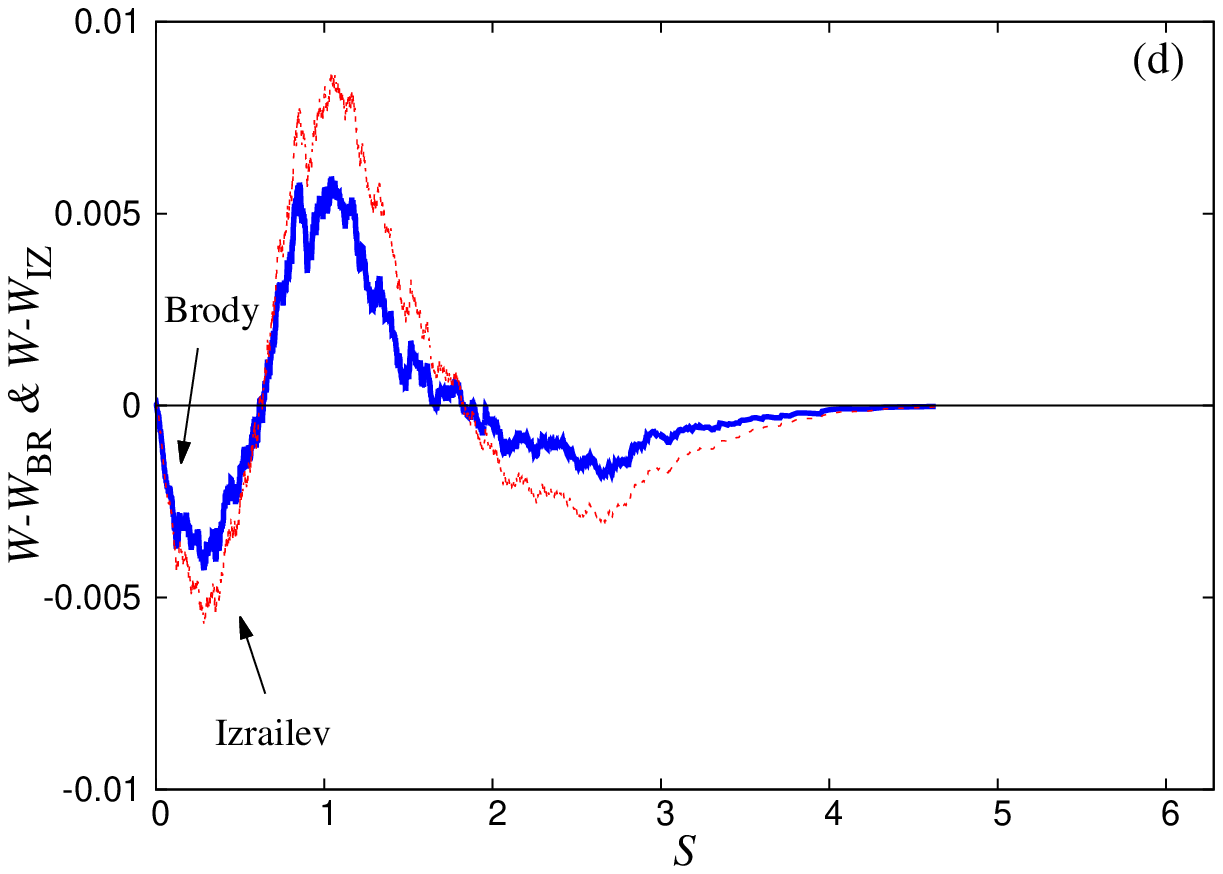}
\caption{Same as Fig.~\ref{fig7} for $M\times N=161\times 398$, $K=7$ and $k=17$ with $\bro=0.67$. See also Table~\ref{tb1} and text for discussion.}
\label{fig9}
\end{figure*}

\begin{figure*}
\center
\includegraphics[width=6cm]{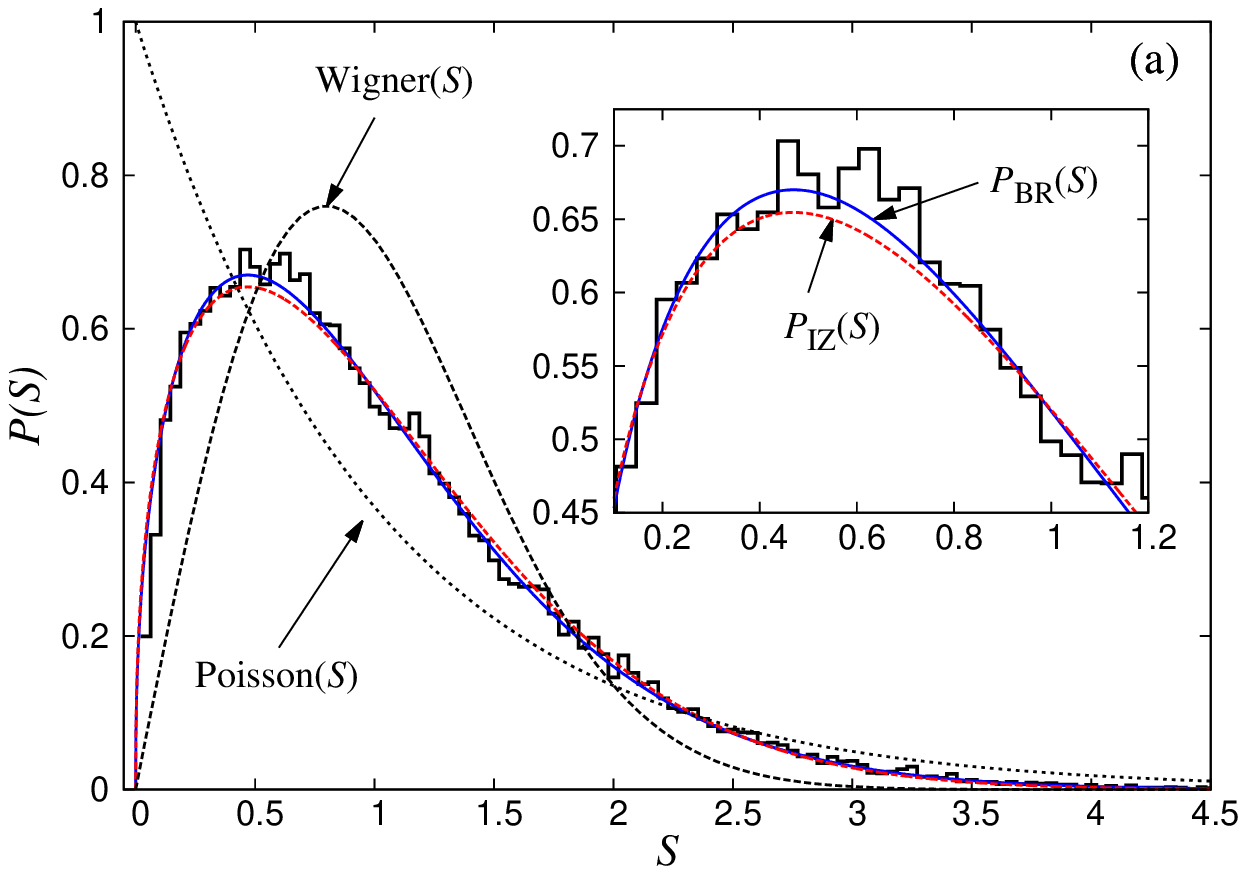}
\includegraphics[width=6cm]{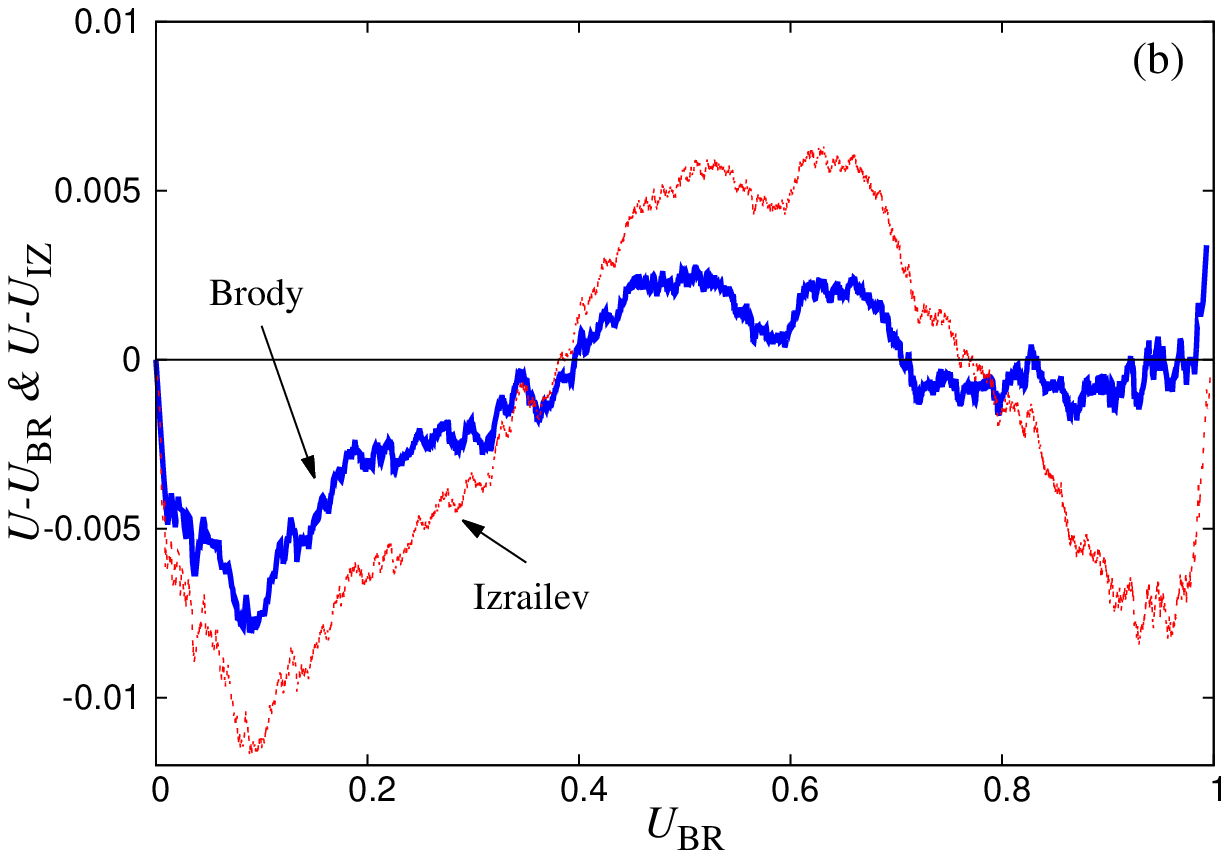}\\
\includegraphics[width=6cm]{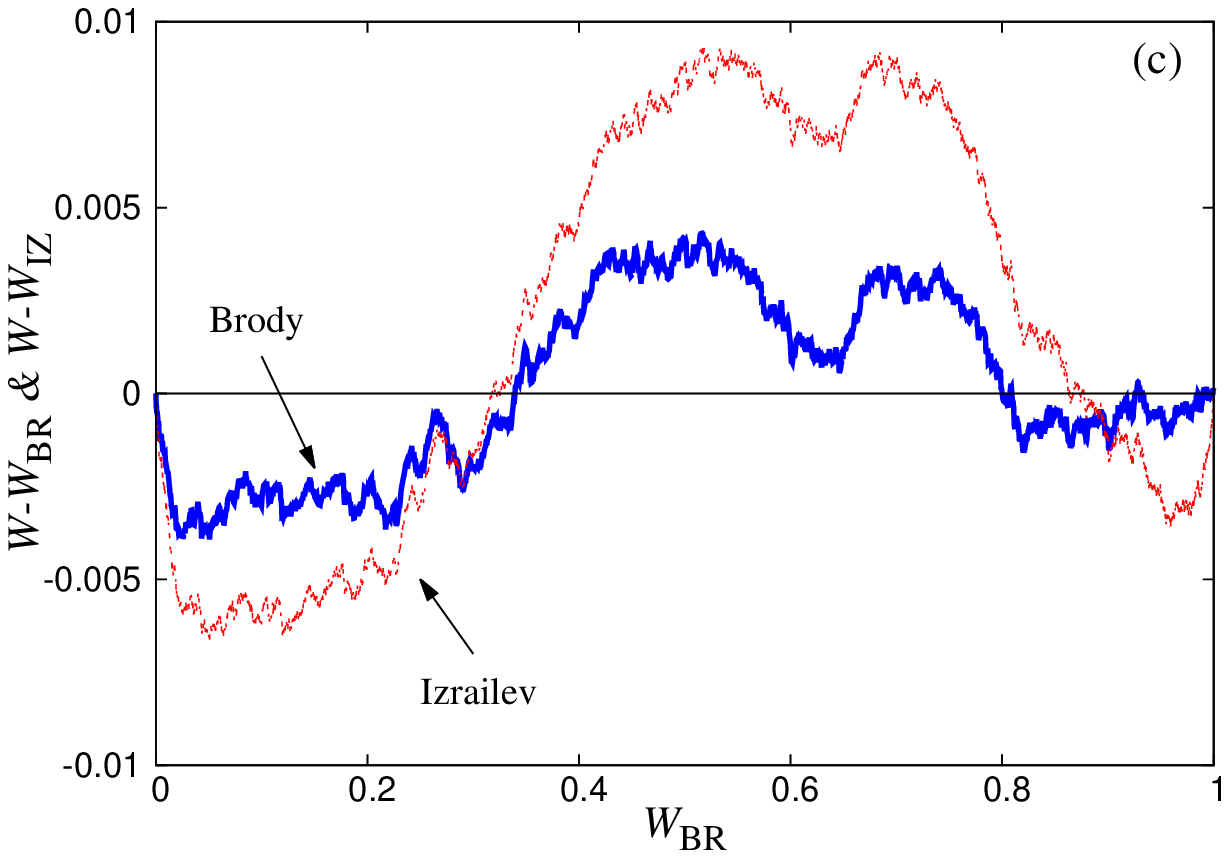}
\includegraphics[width=6cm]{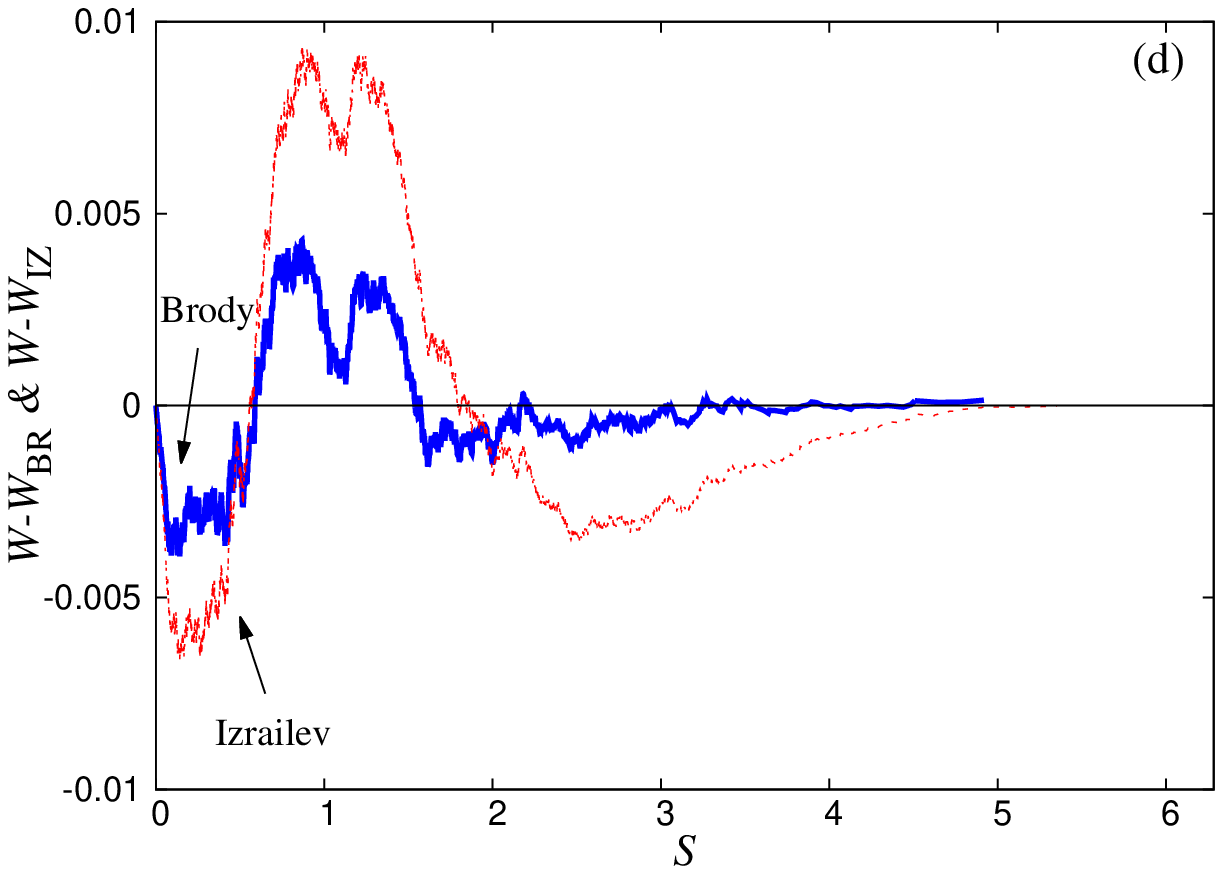}
\caption{Same as Fig.~\ref{fig7} for $M\times N=9\times 4000$, $K=7$ and $k=30$ with $\bro=0.42$. See also Table~\ref{tb3} and text for discussion.}
\label{fig6}
\end{figure*}

In continuation of the section \ref{sec4} we show some more examples of
the statistical spectral analysis for various values of $\bro$ in Figs.~\ref{fig7}, \ref{fig8} and  \ref{fig9}. Finally, we do a similar analysis for large matrices $N=4000$, and take $M=9$ of them in an ensemble. The results are shown in Fig.~\ref{fig6} and more data  are collected in Table~\ref{tb3} which are partially presented in \cite{ManRobBrusConf:2012}. We clearly see that Brody distribution is an excellent fit to the level spacing distribution in all three representations, $P(S)$, $W(S)$ and $U(S)$.

\bibliography{ManRobPRER}

\end{document}